\documentclass{aa}  

\usepackage{graphicx}
\usepackage{txfonts}
\usepackage{psfrag,color,ulem,gensymb}
\definecolor{deepblue}{rgb}{0,0,0.5}
\definecolor{deepred}{rgb}{0.6,0,0}
\definecolor{deepgreen}{rgb}{0,0.5,0}
\definecolor{darkgreen}{rgb}{0,0.6,0}

\usepackage{textcomp}
\usepackage{xspace}
\usepackage{booktabs}
\usepackage[flushleft]{threeparttable}

\newcommand{\tsq}{\hbox{$t^2$}}

\newcommand{\ned}{\hbox{$n_e$}} 
\newcommand{\temp}{\hbox{T$_e$}} 
 
\newcommand{\teq}{\hbox{$T_{eq}$}}

\newcommand{\tmean}{\hbox{$\bar T_0$}}
\newcommand{\Trec}{\hbox{$\bar T_{rec}$}}

\newcommand{\rbh}{\hbox{$r_{BH}$}}
\newcommand{\gamh}{\hbox{$\Gamma_{heat}$}}

\newcommand{\remi}{$r_{{emi}}$}

\newcommand{\rdust}{$r_{\mbox{\scriptsize dust}}$}

\newcounter{{ccc}}

\newcommand{\ramoiii}{\hbox{$\xi_{\rm OIII}^{S_{0}}$}}

\newcommand{\ram}{\hbox{$\xi$}}

\newcommand{\lam}{\hbox{$\lambda$}}
\newcommand{\lamf}{\hbox{$\footnotesize{\lambda}$}}

\newcommand{\nout}{\hbox{\it n$_{H}^{\rm o}$}}

\newcommand{\navr}{\hbox{\it $\bar {n}_{\rm wei}$}}
\newcommand{\ncut}{\hbox{\it n$_{\rm cut}$}}
\newcommand{\nsng}{\hbox{\it n$_{\rm sng}$}}
\newcommand{\nlow}{\hbox{\it n$_{\rm low}$}}

\newcommand{\delt}{\hbox{\it $\Delta {\rm T}_{iso}$}}

\newcommand{\Tosng}{\hbox{\rm T$_{\rm OIII}^{\,sng}$}}
\newcommand{\Topla}{\hbox{${\rm \bar{T}_{\rm OIII}^{\,{PL}}}$}}
\newcommand{\Toplav}{\hbox{$\left<\Topla\right>$}}
\newcommand{\Tosngav}{\hbox{$\left<{\rm T}_{\rm OIII}^{\,sng}\right>$}}

\newcommand{\uout}{\hbox{\it {U}$_{\rm o}$}}

\newcommand{\fmb}{\hbox{\it F$_{MB}$}}
\newcommand{\Tcut}{\hbox{\it T$_{cut}$}}

\newcommand{\uav}{\hbox{$\bar U$}} 
\newcommand{\Av}{\hbox{$A_V$}} 

\newcommand{\Laasto}{\hbox{La$^{1}_{\star}$}}
\newcommand{\Laastd}{\hbox{La$^{2}_{\star}$}}
\newcommand{\Laasav}{\hbox{La$^{3}_{\star}$}}

\newcommand{\rpk}{\hbox{$r_{21}^{pk2}$}}

\newcommand{\Fpko}{\hbox{$F_{La12}^{pk1}$}}
\newcommand{\Fpkt}{\hbox{$F_{*}^{pk2}$}}

\newcommand{\mic}{\hbox{$\mu{\rm m}$}} 
\newcommand{\cms}{\hbox{${\rm cm^{-2}}$}} 
\newcommand{\cmc}{\hbox{${\rm cm^{-3}}$}}

\newcommand{\ami}{\hbox{$A_{M/I}$}}
\newcommand{\Cmi}{\hbox{$C_{M/I}$}}
\newcommand{\Rseq}{\hbox{$R^{i}_{seq}$}}
\newcommand{\Rmb}{\hbox{$R^{i}_{MB}$}}
\newcommand{\Rib}{\hbox{$R^{i}_{IB}$}}

\newcommand{\ev}{\hbox{\sl eV}}

\newcommand{\epsi}{\hbox{$\epsilon$}}

\newcommand{\afir}{\hbox{$\alpha_{25-60}$}}

\newcommand{\aox}{\hbox{$\alpha_{OX}$}}

\newcommand{\auv}{\hbox{$\alpha_{UV}$}}
\newcommand{\afuv}{\hbox{$\alpha_{FUV}$}}

\newcommand{\zsol}{\hbox{$Z_{\sun}$}}

\newcommand{\degk}{K}

\newcommand{\farcsec}{\hbox{$^{\prime\prime}$}}

\newcommand{\hi}{\hbox{H\,{\sc i}}}
\newcommand{\hii}{\hbox{H\,{\sc ii}}}

\newcommand{\hep}{\hbox{He$^{\rm +}$}}

\newcommand{\hyp}{\hbox{H$^{\rm +}$}}

\newcommand{\npp}{\hbox{N$^{\rm +2}$}}
\newcommand{\opp}{\hbox{O$^{\rm +2}$}}
\newcommand{\oppp}{\hbox{O$^{\rm +3}$}}
\newcommand{\nso}{\hbox{$^{\rm 1}{S}_{\rm 0}$}}
\newcommand{\ndd}{\hbox{$^{\rm 1}{D}_{\rm 2}$}}
\newcommand{\ddx}{\hbox{$^{\rm 2}{D}$}}
\newcommand{\dpx}{\hbox{$^{\rm 2}{P}$}}

\newcommand{\spp}{\hbox{S$^{\rm +2}$}}
\newcommand{\sppp}{\hbox{S$^{\rm +3}$}}

\newcommand{\ha}{\hbox{H$\alpha$}}

\newcommand{\hb}{\hbox{H$\beta$}}
\newcommand{\hbw}{\hbox{H$\beta$\,$\lambda$4861\AA}}

\newcommand{\hgw}{\hbox{H$\beta$\,$\lambda$4340\AA}}
\newcommand{\hei}{\hbox{He\,{\sc i}}}
\newcommand{\heii}{\hbox{He\,{\sc ii}}}

\newcommand{\heiwar}{\hbox{He\,{\sc i}\,$\lambda$4713\AA}}

\newcommand{\heiwarivb}{\hbox{\lam5876\AA/\lam4740\AA}}
\newcommand{\fblHe}{\hbox{$f^{HeI}_{\rm blend}$}}
\newcommand{\fblNe}{\hbox{$f^{NeIV}_{\rm blend}$}}
\newcommand{\Rneiv}{\hbox{R$_{\rm Ne}$}}
\newcommand{\Rhei}{\hbox{R$_{\rm He}$}}
\newcommand{\RHeAr}{\hbox{R$_{\rm He/Ar}^{}$}}
\newcommand{\RNeAr}{\hbox{R$_{\rm Ne/Ar}^{}$}}

\newcommand{\sii}{\hbox{[S\,{\sc ii}]}}
\newcommand{\siiw}{\hbox{[S\,{\sc ii}]\,\lam\lam6716,31\AA}}
\newcommand{\siirr}{\hbox{\lam6716\AA/\lam6731\AA}}
\newcommand{\siibr}{\hbox{\lam\lam4069,76\AA/\lam\lam6716,31\AA}}
\newcommand{\oiibr}{\hbox{\lam\lam3726,29\AA/\lam\lam7320,31\AA}}
\newcommand{\ariva}{\hbox{\lam4711\AA}}
\newcommand{\arivb}{\hbox{\lam4740\AA}}
\newcommand{\arivr}{\hbox{\lam4711\AA/\lam4740\AA}}

\newcommand{\arivtoo}{\hbox{\lam\lam\lam4711,13,15\AA/\lam4740\AA}}
\newcommand{\arivtop}{\hbox{\lam4711\AA{\tiny +}/\lam4740\AA}}

\newcommand{\arivl}{\hbox{\lam\lam4711,40\AA}}

\newcommand{\neiva}{\hbox{\lam\lam4715\AA}}

\newcommand{\neivb}{\hbox{\lam\lam4725\AA}}

\newcommand{\oiiir}{\hbox{\lam4363\AA/\lam5007\AA}}

\newcommand{\niir}{\hbox{\lam5755\AA/\lam6583\AA}}
\newcommand{\oiiirb}{\hbox{\lam5007\AA/\lam4861\AA}}
\newcommand{\niirb}{\hbox{\lam6583\AA/\lam6563\AA}}

\newcommand{\nii}{\hbox{[N\,{\sc ii}]}}

\newcommand{\ariv}{\hbox{[Ar\,{\sc iv}]}}
\newcommand{\arivp}{\hbox{[Ar\,{\sc iv}]{\tiny +}}}
\newcommand{\neiv}{\hbox{[Ne\,{\sc iv}]}}

\newcommand{\wavhb}{\hbox{(\lam5007\AA/\lam4861\AA)}}

\newcommand{\Roiii}{\hbox{$R_{\rm OIII}$}}

\newcommand{\oiv}{\hbox{[O\,{\sc iv}]}}
\newcommand{\oivw}{\hbox{[O\,{\sc iv}]\,\lam25.89\,$\mu$m}}

\newcommand{\neiii}{\hbox{[Ne\,{\sc iii}]}}

\newcommand{\oiiix}{\hbox{O\,{\sc iii}}}
\newcommand{\oiii}{\hbox{[O\,{\sc iii}]}}
\newcommand{\oiiiw}{\hbox{[O\,{\sc iii}]\,\lam5007\AA}}
\newcommand{\oiiiww}{\hbox{[O\,{\sc iii}]\,\lam\lam4959, 5007\AA}}
\newcommand{\oiiitw}{\hbox{[O\,{\sc iii}]\,\lam4363\AA}}

\newcommand{\oii}{\hbox{[O\,{\sc ii}]}}
\newcommand{\oiiRL}{\hbox{O\,{\sc ii}}}
\newcommand{\oiiVO}{\hbox{O\,{\sc ii}\,V1}}

\newcommand{\oi}{\hbox{[O\,{\sc i}]}}

\newcommand{\feviiw}{\hbox{Fe\,{\sc vii}\,\lam6086\AA}}

\newcommand{\map}{\hbox{{\sc mappings i}}}
\newcommand{\maphead}{\hbox{{MAPPINGS\,Ig}}}
\newcommand{\ifla}{\hbox{{\sc i}g}}
\newcommand{\MAP}{\hbox{{\sc mappings i}}}
\newcommand{\OSALD}{{\sc osald}}

\begin{document}

   \title{Constraints on the densities and temperature of Seyfert\,2 NLR} 


   \author{Luc Binette
          \inst{1,2}
          \and
          Henry R. M. Zovaro\inst{3,4}
          \and 
          Montserrat Villar Mart\'in\inst{5} 
          \and
          Oli L. Dors\inst{6}
          \and
           Yair Krongold\inst{1}
          \and
          Christophe Morisset\inst{7, 8}
         \and 
          Mitchell Revalski\inst{9}
          \and
         Alexandre Alarie\inst{2}
          \and
          Rogemar A. Riffel\inst{10}
          \and
         Mike Dopita\inst{3}{\textdagger}\fnmsep\thanks{{\textdagger}Deceased }
          }

   \institute{
  Instituto de Astronom\'{\i}a, Universidad Nacional Aut\'onoma de M\'exico, A.P. 70-264, 04510 M\'exico, D.F., M\'exico, M\'exico.
  \and   
  D\'epartement de physique, de g\'enie physique et d'optique,  Universit\'e Laval, Qu\'ebec, QC G1V 0A6, Canada. 
  \and
  Research School of Astronomy \& Astrophysics, Australian National University, Canberra 2611, Australia
  \and
  ARC Centre of Excellence for All Sky Astrophysics in 3 Dimensions (ASTRO3D), Australia
   \and
  Centro de Astrobiolog\'\i{}a, (CAB, CSIC-INTA), Departamento de Astrof\'\i{}sica, Cra. de Ajalvir Km. 4, 28850, Torrej\'on de Ardoz, Madrid, Spain
  \and  
  Universidade do Vale do Para\'{\i}ba. Av. Shishima Hifumi, 2911, CEP: 12244-000, São Jos\'{e} dos Campos, SP, Brazil 
   \and
   Universidad Nacional Autónoma de México, Instituto de Astronomía, AP 106,  Ensenada 22800, BC, México
  \and
  Instituto de Ciencias Físicas, Universidad Nacional Autónoma de México, Av. Universidad s/n, 62210 Cuernavaca, Mor., México
 \and
  Space Telescope Science Institute, 3700 San Martin Drive, Baltimore, MD 21218, USA
  \and
  Departamento de F\'isica, Centro de Ci\^encias Naturais e Exatas, Universidade Federal de Santa Maria, 97105-900, Santa Maria, RS, Brazil 
      }               

   \date{Received 23, 12, 2022; accepted 02, 01, 2024}

  \abstract
   {Different studies have reported the so-called temperature problem of the narrow line region (NLR) of active galactic nuclei (AGNs). Its origin is still an open issue. To properly address its cause, a trustworthy temperature indicator is required. }  
   {No space for them in arXiv}
   {We propose that the weak \ariv\ $\lambda\lambda$4711,40\AA\ doublet is the appropriate tool for evaluating the density of the high excitation plasma. We subsequently made use of the recent S7 survey sample to extract reliable measurements of the weak \ariv\ doublet in 16 high excitation Seyfert\,2s. As a result we could derive the plasma density of the NLR of our Seyfert\,2 sample and compared the temperature inferred from the observed \oiii\ (\oiiir) ratios. }   
   {It was found that 13 Seyfert\,2s cluster near similar values as the \oiii\ (\oiiir)  ratio, at a mean value of 0.0146$\,\pm\,$0.0020. Three objects labeled outliers stand out at markedly higher \oiii\ values ($>0.03)$.
   }    
   {If for each object one assumes a single density, the values inferred from the \ariv\ doublet for the 13 clustering objects all lie below $60\,000\,$\cmc, indicating that the \oiii\  (\oiiir) ratios in these objects is a valid tracer of plasma temperature. Even when assuming a continuous power-law distribution of the density, the inferred cut-off density required to reproduce the observed \ariv\ doublet is in all cases $< 10^{5.1}$\,\cmc. The average NLR temperature inferred for the 13 Seyfert\,2s is $13\,000\pm703$\,\degk, which photoionization models have difficulty reproducing. Subsequently we considered different mechanisms to account for the observed \oiii\ ratios. For the three outliers, a double-bump density distribution is likely required, with the densest component having a density $> 10^6\,$\cmc.  }  
   
\keywords{Galaxies: emission lines: Seyfert  -- line: formation -- plasmas: emission lines -- active galaxies}

   \maketitle
%

\section{Introduction}
\label{sec:intro}

Prevailing models of the narrow-line region (NLR) of
active galactic nuclei (AGNs) consider a distribution of photoionized clouds that extends over a wide range of cloud densities and ionization parameter values, whether the targets are Type\,I AGNs \citep{Ba95,Kor97a,BL05} such as quasars\footnote{The term "quasar" is used to refer to Type\,I AGNs.} (QSO\,1s), Seyfert\,1s and broad-line radio galaxies, or alternatively Type\,II objects \citep[]{Fg97,Ri14} which consist of Seyfert\,2s, QSO\,2s and narrow-line radio galaxies (NLRGs). 

A concern when modeling the AGNs emission lines is the handling of the plasma density, a parameter that can impact forbidden line ratios and affect temperature estimates. AGN-driven outflows are believed to play a crucial role in galaxy evolution by curtailing star formation in massive galaxies, thereby suppressing the high-mass end of the galaxy luminosity function~\citep[e.g.,][]{Croton2006}.
Accurately estimating outflow mass rates and energetics is therefore critical in understanding precisely how these outflows regulate star formation across cosmic time.  Constraining these parameters requires knowledge of the plasma density gradient toward the nucleus, as this is needed to measure the ionized gas mass, which is in turn essential in evaluating mass outflow rates and wind energetics~\citep[e.g.,][]{Nesvadba2008}. 

At densities above the critical density $n_{\rm crit}$ of a particular forbidden line transition, collisional deexcitations from the excited states begin to dominate over radiative deexcitations, reducing the strength of the corresponding emission lines \citep{OST89}. In the particular case of the \oiii\  (\oiiir) ratio (hereafter \Roiii), in order to reliably determine the plasma temperature of the high ionization emission regions, we must quantify to which extent this ratio is affected or not by collisional deexcitation. For this purpose,  we rely on the weak \ariv\ $\lambda\lambda$4711,40\AA\ doublet, which is a  "direct" density indicator appropriate to high ionization plasmas \citep[][]{Wa04,Ke19}.

This work is a follow up of the \citet[][hereafter BVM]{BVM} paper which studied the Seyfert\,2 sample of \citet[hereafter Kos78]{Kos78}, which had the particularity of providing reliable  measurements of the \ariv\ doublet in seven objects. The average NLR temperature BVM inferred from \Roiii\ assuming a power-law density distribution was 13\,500\,\degk, which standard single-density photoionization models cannot reproduce, underscoring the so-called "temperature problem" (hereafter TE\,problem), that is, where single-zone low density photoionization models underpredict the observed \Roiii\ ratios, as reported in various NLR studies \citep[][]{SB96,Be06a,VM08,Dr15}. In the current work, we determined the NLR densities and corresponding plasma temperatures using a larger sample consisting of 16 Seyfert\,2s from the S7 survey \citep{Do15, TA17}, which present reliable measurements of the weak \ariv\ doublet. We subsequently explored possible explanations to account for the larger temperatures observed with respect to standard photoionization calculations.

Following the introduction of Sect.\,1,  we compare in Sect.\,2  various methods for determining the NLR densities and in particular the advantage of using the \ariv\ \lam4711,40\AA\ doublet lines. In Sect.\,3 we describe the S7 survey and the procedure adopted for the extraction of the \ariv\ doublet while in Sect.\,4, using the \oiii\ (\oiiir) lines, we determine the temperature characterizing the Seyfert\,2 galaxies of our sample taking into account the densities we inferred from the \ariv\ doublet. In Sect.\,5 we look at the implication of the densities encountered in the determination of outflow kinematics.  The plasma temperatures we derive are higher than those predicted by the standard photoionization models as indicated in Sect.\,6, while in Sect.\,7 we review alternative explanations such as a double-bump ionizing energy distribution, the predominance of matter-bounded photoionized components, the presence of temperature fluctuations, and finally the possibility of a non-Maxwellian electron energy distribution. Throughout this paper, the adopted cosmological parameters are
$\rm H_{0}=67.4 \: \rm km \:s^{-1} Mpc^{-1}$ and $\Omega_{\rm m}=0.315$ \citep{Pl652}.

\section{Studies of the NLR and ENLR of active galactic nuclei}
\label{sec:forbi}

\subsection{Contrast between Type\,I and II narrow line region}
\label{sec:contrast}

In the NLR of Type\,I AGNs, significant collisional deexcitation of the   \oiiiww\  lines\footnote{From level $^1\rm{D}_2$ to levels $^3\rm{P_1}$ and $^3\rm{P_2}$.} takes place, as corroborated by the work of \citet[hereafter BL05]{BL05} who compared the \Roiii\ (\oiiir) ratios observed in 30 quasars from the \citet{BG82} sample. The measured ratios extended from 0.02 to 0.2. On the other hand, as pointed out by BVM, the Type\,II AGNs from the Kos78 sample show \Roiii\ similar to those observed in the spatially resolved extended NLR (hereafter ENLR) where the densities encountered are moderate, typically $< 10^3\,$\cmc. Although some level of collisional deexcitation might be present in the Kos78 sample, BVM concluded that its impact on the observed \oiii\ lines fluxes was not significant given that the densities they inferred from the \ariv\ doublet were all below $10^4\,$\cmc. The explanation they proposed for their Seyfert\,2s to share a similar \Roiii\ ratio that does not extend to the higher values commonly observed in quasars is the orientation of the AGN ionizing cone with respect to the observer. 

As suggested earlier on by \citet[hereafter NMT]{NMT}, the geometrical set up behind the AGN unified model \citep{An93} might apply not only to the BLR but to the inner denser parts of the NLR which possibly becomes partly obscured in Type\,II objects. 
NMT for instance found that Type\,I Seyferts show a statistically higher \Roiii\ than Type II Seyferts. \citet[]{Me08a} favor a similar interpretation with respect to the mid-infrared coronal lines by pointing out that the mean \oiiiw\ line luminosity is 1.4\,dex smaller in Seyfert\,2s than in Seyfert\,1s while in the case of the mean luminosity of the far-infrared \oivw\ line, the difference between the two subgroups is only 0.2\,dex. This corroborates the earlier works of \citet[][]{JB91,CAM93,Mu94,Ke94,RL05,Nz06} who pointed out that a much higher dust extinction affects the NLR of Seyfert\,2s in comparison to Seyfert\,1s. Interestingly, \citet{Kr11} found that among AGNs with strong inclinations (i.e., with $b/a > 0.5$), a comparison of the \Roiii\ and \oiii/\oiv\ (\lam5007\AA/28.59$\mu$m) ratios among their sample reveal that Seyfert 2's tend to present lower values of both ratios with respect to Seyfert\,1s, which they proposed is indicative that more extinction of the NLR emission is present in Seyfert\,2s.

The ENLR densities inferred from the red \siiw\ doublet by \citet[][]{Be06a,Be06b} appear in most cases  to be increasing toward the nucleus. If this gradient extended inward toward the supermassive black hole (SMBH), the resulting densest NLR sections might be expected to become hidden in Seyfert\,2s although not in Seyfert\,1s. A graphical description of such geometry is illustrated in Fig.\,2 of \citet{Be06c}. Further information concerning the plasma densities inferred from the \sii\ doublet is given in App.\,\ref{sec:ap-sii}.

\subsection{Densities characterizing the spatially resolved ENLR}
\label{sec:introa}

By ENLR we emphasize that we specifically refer to emission from plasma components that are spatially resolved. For the Seyfert\,2s of our sample, this implies any plasma localized at a projected radial distance \rbh\  (from the SMBH) higher than 1.5 times the seeing size. In brighter objects such as QSO\,2s, however, this minimum distance has to be larger since the central source can dominate the emission up to several times the seeing size due to atmospheric blurring, causing a mixing of the unresolved nucleus emission with the foreground ENLR emission \citep{VM16}.  For observations taken with the \textit{Hubble} Space Telescope (HST), since no seeing is present, the ENLR initiates closer to the SMBH, at a projected radius \rbh\ equals to the slit width. Incidentally, ground-based adaptive optics measurements such as obtained with VLT MUSE offer a spatial resolution approaching that of HST \citep[e.g.,][]{Wi22}. 

With HST observations, owing to the high spatial resolution, the ENLR can be resolved down to a smaller radial distance  of typically $\simeq 0.2$\farcsec\ from the SMBH.  For instance, while ground-based observations of \Roiii\ of the nucleus of NGC\,1068 indicate a value comparable to that observed in other Seyferts\,2s (BVM), the  HST-FOS measurements analyzed by \citet{Kr98} show emission from two  knots situated at distances \rbh\ from the nucleus of 0.7\farcsec\ and 0.2\farcsec, respectively (i.e., at  57 and 16\,pc, respectively), which are superposed to a diffuse background emission. The spectra of these knots are consistent with a radial increase  of \Roiii\ toward the spatially unresolved nucleus where it reaches its highest value (see position of the three knot \Roiii\  measurements in Fig.\,1 of BVM). The diffuse plasma emission surrounding the knots and the nucleus presumably dominates in luminosity since ground-based observations of the nuclear region do not show an unusual integrated \Roiii\ ratio. The  gradient in the \Roiii\ ratio across the knots and the unresolved inner nucleus might be the result of an increase in temperature due to shock excitation, or more likely, to collisional deexcitation, as was proposed by BL05 for the NLR of their quasar sample. 

To confirm when collisional deexcitation of \oiii\ is present or not, a reliable option to evaluate the densities consists in using the  \ariv\ (\arivr) doublet ratio. Using the Magellan Echellette spectrograph mounted on one of the Magellan Telescope,  \citet{Co17} succeeded in measuring the radial behavior of the \ariv\ doublet ratio within the ENLR of two Seyfert\,2s, IC\,5063 and NGC\,7212. The densities they inferred were all below $10^4\,$\cmc, except at the nucleus on the north side of IC\,5063. Using the VLT Xshooter spectrograph, \citet[][]{Ho23} obtained deep observations of IC\,5063 where the \ariv\ doublet measured in the nucleus implied  a  density in $\log_{10}$ of ${3.79}^{+0.18}_{-0.19}\,\cmc$.

\subsection{Prior studies of NLR densities in Seyfert\,2s }
\label{sec:tradi}
 
Evidence of collisional deexcitation of \Roiii\ ratio is presented by \citet{KS97} who addressed the TE\,problem by exploring a wide range of parameters in their photoionization calculations. Five of the  36  Seyfert\,2 of their sample indicated \Roiii\ ratios above 0.03.  Single density matter-bounded models with $\ned \ga 10^{5}\,$\cmc\ and a high ionization parameter could reproduce the observed high \Roiii\ ratios. However, in order to reproduce the line ratios of other ionic species (\nii, \oi, \oii, \neiii\, etc), the authors had to adopt a multicomponent approach that combines photoionization models of four different densities ($\log n= 2,3,4, 5$) independently distributed at four galactocentric radii. These components shared the same weight in \hb\ luminosity (30\%), except the densest (10\% for $\log n= 5$). In conclusion, the models that \citet{KS97} favored assume subsolar metallicities and an ionizing SED consisting of a thermal bump superposed to a steep $\auv=-2$ power law. This procedure lead to a reasonable match to the behavior of the selected line ratios although the highest observed values of $\Roiii \approx 0.03$ could not be reproduced.   

Another version of multicomponent calculations consists of the Locally Optimally emitting Clouds (LOC) models of \citet[hereafter Ri14]{Ri14} and \citet[hereafter Fg97]{Fg97} who calculated an extensive grid of models where each calculation integrates the line fluxes over a wide range of both radii from the ionizing source and of cloud densities at each radius (up to $10^{10}\,$\cmc). The abundances adopted corresponded to 1.4\,\zsol. To model the  high excitation Seyfert\,2s, their target ratios  consisted of the  subset a41, which represents a weighted average of different optical lines derived from the Sloan Digital Sky survey  (SDSS) survey by {Ri14} after applying the Mean Field Independent Component Analysis (MFICA) tool \citep[see also][]{Allen2013} to their SDSS sample consisting of $\sim 10^4$ Type\,II emission line galaxies in the redshift range $0.1 < z < 0.12$. Given the wide distribution of the densities covered by LOC models, collisional deexcitation is fully taken into account and it allowed their models to reproduce the \Roiii\ ratio of the a41 subset as well as the bulk of the other line ratios considered.  As pointed out by BVM, however, no direct evidence was found that the densities of the observed NLR in Seyfert\,2s extended beyond $10^5\,$\cmc, even after assuming a power-law density distribution that extended up to a density cut-off consistent with the observed \ariv\ doublet ratio, which suggests that the much denser NLR observed in Type\,I AGNs (see BL05) might generally be hidden in Seyfert\,2s. In the current study we apply a similar approach but to a larger sample.  

\subsection{An approach based on the \ariv\ \arivl\ diagnostic}
\label{sec:adoubl}

The \ariv\ doublet is a direct density diagnostic well suited to studies of high excitation plasma \citep[see][]{Ke19}. It has  traditionally been  used in the context of planetary nebulae (PNe). Since the ion $\rm Ar^{+3}$ is located in the high ionization plasma, the densities inferred are more directly related to the \oiii\ emission regions than those derived from the \sii\ red doublet, which originates from the low ionization plasma. Its relative weakness, however, has hampered its use in AGNs studies.   
For instance, out of the 2153 Seyfert\,2s extracted from the SDSS-DR7 data release by \citet[]{Va12}, the \ariv\ doublet could be measured in only seven objects while the \oiiitw\ line was measured in 86 objects and the \feviiw\ line in 96 objects. Interestingly, the densities inferred by \citet[]{Va12} from the \ariv\ doublet assuming a temperature of $10^4\,$\degk\ were on the order of $10^3\,$\cmc\ (for their seven Seyfert\,2s) while those they inferred using the  \sii\ doublet in 2300 objects showed a median value of 250\,\cmc\ and only in 97 objects were the densities found to be higher than $10^3\,$\cmc. The aim of the current study is to extend  to a larger sample the analysis of BVM who disposed of only seven high excitation Seyfert\,2s with measured \ariv\ (\arivr) ratios. 

Assuming a 14\,000\,\degk\ plasma,  the critical densities\footnote{In accordance to the definition of critical density from \citet{OST89}.}  for the upper levels of the \ariv\ \lam4711\AA\ and \lam4740\AA\ lines are $1.69 \times 10^4$ and $1.52 \times 10^{5}$\,\cmc\ while for the \Roiii\ temperature indicator, the critical densities of the upper levels of the \oiii\ \lam5007\AA\ and \lam4363\AA\ lines are $7.67\times 10^5$ and $2.79\times 10^7$\,\cmc, respectively.  These results were derived using the {\sc PyNeb} library version 1.1.18 \citep{luridianaetal15}. The atomic data for \ariv\ used by default in {\sc PyNeb} are taken from \citet{2019Rynkun_aap623} and \citet{RBK97}, and for \oiii\ from \citet{SSB14}, \citet{2000Storey_mnra312} and \citet{2004Froese-Fischer_Atom87}. The effect of atomic data on the densities determined from a given diagnostic line ratio can be important \citep{Morisset20}. We have explored all the combinations of atomic data for \ariv\ published after 1980, which are available in {\sc PyNeb},  and obtain very similar values. The same applies for the critical densities obtained for \oiii\ where very similar results are also obtained, regardless of the atomic data used.

\section{The Seyfert\,2 sample extracted from the S7 survey}
\label{sec:obs}

\begin{figure*}  %
\centering
\includegraphics[width=17cm]{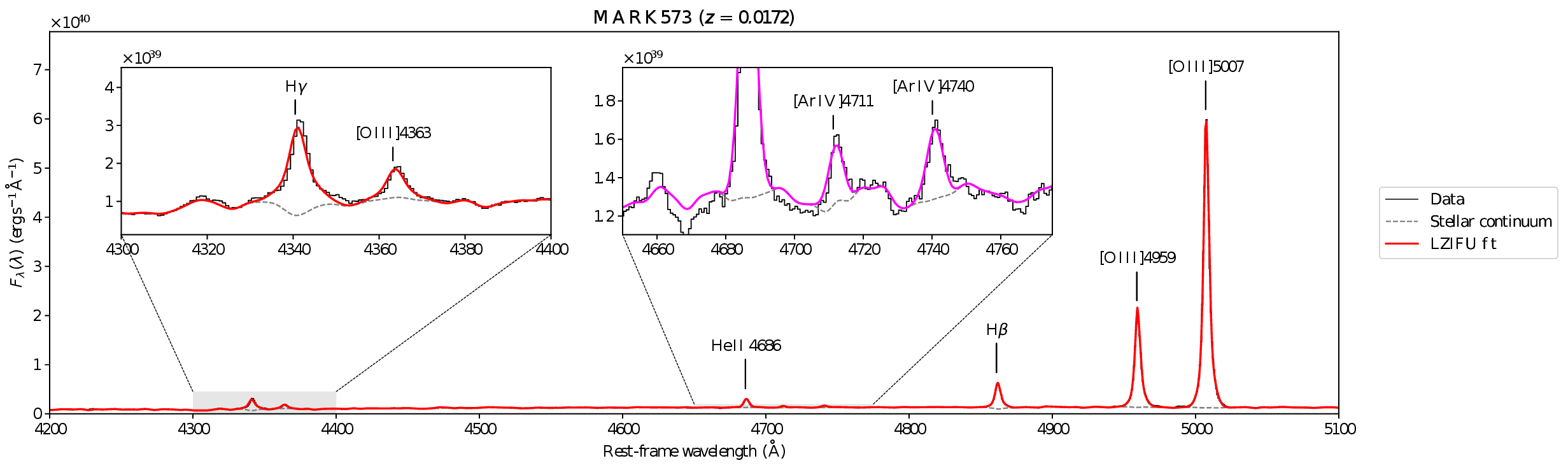}
\caption{Observed spectrum of Mrk\,573 (black solid line). The \textsc{lzifu} emission line fit (red) is superimposed to the dashed line which represents the stellar continuum fit.
}
\label{fig:fig1}
\end{figure*}

\subsection{The S7 survey data release}
\label{sec:data}

Our sample was originally extracted from the final data release \citep{Do15, TA17} of the Siding Spring Southern Seyfert Spectroscopic Snapshot Survey\footnote{Currently available at \url{https://docs.datacentral.org.au/s7/}}. The S7 survey was performed over 2013–2016 using the Wide Field Spectrograph (WiFeS) on the ANU 2.3\,m telescope at Siding Spring Observatory \citep{Do15}. The sample consists of 131 galaxies where most have been selected from the \citet{VC06,VC10} catalogs of AGNs.  

Among the 122 galaxies of the S7 sample with reliable measurements of the \oiiitw\ line, we excluded objects classified as Seyfert\,1s, or starbursts or having hybrid nuclei. As in the BVM study, we only considered objects where the \ariv\ (\arivr) ratio is satisfactorily measured. This allows us to evaluate to what extent the optical \oiii\ emission lines  are affected by collisional deexcitation, which is frequently observed to be the case in the NLR of quasars \citep{BL05}. We only considered high excitation Seyfert\,2s, that is objects with a dereddened \oiii/\hb\ \wavhb\ ratio  $> 7.0$. It was not possible to define an excitation sequence of Seyfert\,2 spectra since there were too few objects of lower excitation with a detectable \ariv\ doublet. One advantage is that we are avoiding possible degeneracies due to a mixture of ionization mechanisms and/or to a wide range of ionization parameter, $U$.  
Our sample represents objects of comparable excitation that likely share the same excitation mechanism, that is photoionization from the AGN nuclear accretion disk. 

Our final sample consists of 16 high excitation objects as listed in Table~\ref{tab:sam}. Their redshift covers the range $0.004\le\,z\le\,0.0171$.  After we excluded the three outliers (labelled a,b,c) with anomalous \Roiii\ ratios (as discussed in Sect.\,\ref{sec:outl}), the average \oiii/\hb\ (\lam5007\AA/\lam4861\AA) ratio is $9.67$, which is comparable to the high excitation subset a41 from \citet{Ri14} with $\oiii/\hb=8.6$. 

\subsection{New emission line extraction }
\label{sec:lifit} 

The flux of the blended \arivp\ \ariva\ line\footnote{The appended {\tiny +} symbol denotes the presence of line blending.} is on average three times weaker than the \oiii\ \lam4363\AA\ line; as a result, accurate density estimates using the \ariv\ doublet ratio require high S/N spectra.
Visual inspection of the \ariv\ fits provided in the official S7 data release showed them to be unreliable in a large fraction of galaxies due to their faintness and prominent stellar absorption features. We therefore opted to use a bespoke emission line fitting procedure to obtain more accurate fluxes. 

\subsubsection{Emission line fitting procedure }
\label{sec:procedure}

Most galaxies in our sample exhibit complex emission line profiles due to AGN-driven winds and outflows, or strong jet-ISM interactions, as in the case of IC\,5063~\citep{Mo15}. As a result, multi-Gaussian component fits are required to accurately capture the fluxes of strong lines such as \oiiiw\ and \hb. However, fitting multiple Gaussians that target the much weaker blended lines such as \hei\ \lam4713\AA\ or \neiv\ \neiva\ cannot be justified at the comparatively low S/N of these components. Furthermore, this could introduce systematic errors due to possible mismatch of the stellar templates that are being fit. For this reason we adopted the following consistent approach to derive accurate measurements of the emission line fluxes.

First, the blue and red datacubes were spliced together to create a single "combined" datacube spanning 3500\,--7000\,\AA. This was achieved by spectrally convolving the higher-resolution R7000 cube with a Gaussian kernel in wavelength to the same resolution as the B3000 cube before splicing them together.

The aperture spectra were extracted from the B3000 and combined data cubes using the galaxy center coordinates given in table \texttt{S7\_DR2\_Table\_2\_Catalogue} except for 12 galaxies\footnote{NGC4\,24, NGC\,1052, ESO\,362-G18, MCG-01-24-012, NGC\,3281, NGC3\,783, NGC\,3831, NGC\,4593, ESO\,339-G11, NGC\,6860, IC\,5063 and NGC\,7590.} for which visual inspection showed the supplied values to be off-center. Center coordinates for these galaxies were taken to be the brightest pixel in a white-light image computed by collapsing the combined cube in wavelength. Spectra were corrected for foreground Galactic extinction using the dust map of \citet{SF11} together with the \citet{FM07} extinction curve with $R(V) = A_V/E(B-V)= 3.1$.

\begin{table*}
\caption{High excitation Seyfert\,2s from the S7 survey with reliable measurements of the \ariv\ doublet}
\label{tab:sam}
\small
\begin{tabular}{clccccc}
\toprule
 (1)  & \hspace{0.5cm} (2) & (3) & (4) & (5) & (6)\tablefootmark{a} & (7)\tablefootmark{a}  \\ 
  \cmidrule{1-7}
Index & Object Name & Aperture & Redshift  & \Av & \oiii/\hb & \heii/\hb \\ 
\#  &  &  arcsec & $z$ &  & $\frac{\lamf5007}{\lamf4861}$ & $\frac{\lamf4686}{\lamf4861}$  \\ 
\cmidrule{1-7}

1 & NGC\,5506 & 1" & 0.0061 & 2.85 & 7.08$\,\pm\,$0.10 & 0.15$\,\pm\,$0.014 \\  
2 & NGC\,1320 & 1" & 0.0093 & 1.84 & 9.43$\,\pm\,$0.11 & 0.34$\,\pm\,$0.012 \\
3 & NGC\,5643 & 1" & 0.0040 & 0.00 & 11.1$\,\pm\,$0.11 & 0.23$\,\pm\,$0.009 \\  
4 & Mrk\,573 & 4" & 0.0172 & 1.11 & 10.2$\,\pm\,$0.04  & 0.30$\,\pm\,$0.004 \\  
5 & NGC\,3281 & 4" & 0.0107 & 1.72 & 8.19$\,\pm\,$0.08 & 0.33$\,\pm\,$0.007 \\  
6 & ESO\,137-G34 & 4" & 0.0091 & 1.84 & 9.56$\,\pm\,$0.05 & 0.22$\,\pm\,$0.005\\
7 & NGC\,4939 & 4" & 0.0104 & 1.62 & 10.2$\,\pm\,$0.06 & 0.30$\,\pm\,$0.006 \\  
8 & NGC\,3393 & 1" & 0.0125 & 1.16 & 9.04$\,\pm\,$0.07 & 0.24$\,\pm\,$0.006 \\  
9 & NGC\,7682 & 1" & 0.0171 & 1.30 & 9.75$\,\pm\,$0.11 & 0.18$\,\pm\,$0.010 \\  
10 & IC\,5063  & 4" & 0.0114 & 2.39 & 8.47$\,\pm\,$0.05 & 0.11$\,\pm\,$0.004 \\  
11 & NGC\,5728 & 4" & 0.0093 & 1.21 & 8.70$\,\pm\,$0.06 & 0.18$\,\pm\,$0.007 \\  
12 & IC\,4995 & 1" & 0.0161 & 0.83 & 12.3$\,\pm\,$0.13 & 0.30$\,\pm\,$0.010 \\  
13 & IC\,2560 & 1" & 0.0098 & 0.60 & 11.6$\,\pm\,$0.15 & 0.28$\,\pm\,$0.012 \\  

\cmidrule{1-7}
Outliers  &  &  &  & Averages: & \!\!\!\!\!9.67 & \!\!\!\!\!0.243$\,\pm\,$0.072\\ 
\cmidrule{1-7}
a & ESO\,138-G01 & 4" & 0.0091 & 0.87 & 8.44$\,\pm\,$0.029 & 0.27$\,\pm\,$0.003 \\  
b & Mrk\,1210 & 4" & 0.0135 & 1.00 & 9.38$\,\pm\,$0.040 & 0.20$\,\pm\,$0.004 \\  
c & NGC\,4507 & 4" & 0.0118 & 1.15 & 8.06$\,\pm\,$0.021 & 0.17$\,\pm\,$0.002 \\  
\bottomrule

    \end{tabular}
\tablefoot{
     \tablefoottext{a}{Line ratios were corrected for reddening assuming a Balmer decrement of 3.1. }
}
\end{table*}


Emission lines were fitted using the \textsc{idl} package \textsc{lzifu}\footnote{\url{https://github.com/hoiting/LZIFU}}\citep{Ho2016}. To extract accurate emission line fluxes, \textsc{lzifu} first uses \textsc{ppxf}~\citep{CE04,Ca17} to fit and subtract the stellar continuum by computing the linear combination of template spectra that best fits the data. We used the \citet{Go05} stellar templates due to their high spectral resolution. After subtraction of the stellar continuum, the emission lines are fit simultaneously using one or more Gaussian components, where the line-of-sight velocity and velocity dispersions are constrained to be the same for all emission lines within each component. Optional Legendre polynomials may be included during the stellar continuum and emission line fits to account for stellar template mismatch and/or flux calibration errors. 

The nuclear emission lines were extracted from both the 1\farcsec and 4\farcsec apertures. Although a larger aperture yields higher S/N, it also results in greater contamination from the stellar continuum, which has a variety of complex absorption features in the vicinity of the \ariv\ lines, as illustrated by the spectra of Mrk\,573 in Fig.\,\ref{fig:fig1}. The final aperture chosen for each object was selected based on the balance between emission line S/N and the severity of systematic errors in the stellar continuum fit. 

For red lines fluxes, including \ha\ \lam6563\AA\ and \hei\ \lam5876\AA, we performed one, two and three component fits to the combined spectrum, where the optimal number of components was evaluated using the likelihood ratio test with a critical threshold of 0.01~\citep[e.g.,][]{Ho2016}. A 5th-order additive polynomial was included in the stellar continuum fit to correct for flux calibration errors. We then repeated this process on the B3000 spectrum in order to obtain \oiiiw, \hbw, \hgw\ and \oiiitw\ fluxes. For these fits, a 12th-order additive polynomial was included in the stellar continuum fit to correct for flux calibration errors, and a 12th-order additive polynomial was included during the emission line fit to compensate for residuals due to stellar template match.

\begin{table*}
\caption{NLR temperatures derived using the reddening-corrected ratios for our sample extracted from the S7 survey}
\label{tab:sng}
 \setlength{\tabcolsep}{0.6\tabcolsep}
\small
\begin{tabular}{clccccccccc}
\toprule
 (1)  & \hspace{0.5cm} (2)\tablefootmark{a} & (3) & (4)\tablefootmark{b} & (5) & (6)\tablefootmark{c} & (7)  & (8)\tablefootmark{d} & (9)\tablefootmark{e} & (10)\tablefootmark{f} & (11)\tablefootmark{g} \\
Index & Object Name & \Roiii & \ariv{\tiny +}  & \RHeAr & \fblHe & \neiv/\ariv\ & \fblNe  & \ariv & \nsng & \Tosng \\ 
\#  &  &  $\frac{\lamf4363}{\lamf5007}$ & $\frac{\lamf4711{\tiny +}}{\lamf4740}$  & $\frac{\lamf5876}{\lamf4740}$ &  & $\frac{\lamf\lamf4725}{\lamf4740}$ &  & $\frac{\lamf4711}{\lamf4740}$ &  \cmc  & \degk  \\ 
  \cmidrule{1-11}
1 & NGC\,5506 & 0.0161$\,\pm\,$0.0031 &  1.27$\,\pm\,$0.21 & 5.23$\,\pm\,$0.856 & 0.185 & --  & -- & 1.07 & 2428 & 13\,803 \\
2 & NGC\,1320  & 0.0116$\,\pm\,$0.0012 & 0.85$\,\pm\,$0.08 & 0.84$\,\pm\,$0.090 & 0.040 & -- & -- & 0.82 & 6038 & 12\,073 \\
3 & NGC\,5643  & 0.0117$\,\pm\,$0.0009 &  0.85$\,\pm\,$0.13 & 1.06$\,\pm\,$0.157 & 0.050 & -- & -- & 0.81 & 6194 & 12\,108 \\
4 & Mrk\,573   & 0.0171$\,\pm\,$0.0004 &  0.82$\,\pm\,$0.04 & 1.02$\,\pm\,$0.047 & 0.051 & -- & -- & 0.78 & 7226 & 14\,030 \\
5 & NGC\,3281  & 0.0118$\,\pm\,$0.0016 &  1.08$\,\pm\,$0.10 & 1.30$\,\pm\,$0.120 & 0.068 & 0.45$\,\pm\,$0.30 & 0.38 & 0.75 & 7698 & 12\,115 \\
6 & ESO\,137-G34 & 0.0145$\,\pm\,$0.0007 &  0.79$\,\pm\,$0.10 & 1.84$\,\pm\,$0.151 & 0.099 & -- & -- & 0.72 & 8527 & 13\,086 \\
7 & NGC\,4939  & 0.0146$\,\pm\,$0.0008 &  0.90$\,\pm\,$0.07 & 1.06$\,\pm\,$0.056 & 0.058 & 0.22$\,\pm\,$0.14 & 0.19 & 0.72 & 8705 & 13\,125 \\
8 & NGC\,3393  & 0.0127$\,\pm\,$0.0007 &  0.75$\,\pm\,$0.11 & 1.60$\,\pm\,$0.170 & 0.091 & -- & -- & 0.69 & 9375 & 12\,426 \\
9 & NGC\,7682  & 0.0172$\,\pm\,$0.0011 &  0.71$\,\pm\,$0.12 & 1.47$\,\pm\,$0.169 & 0.088 & -- & -- & 0.65 & 10\,990 & 13\,987 \\
10 & IC\,5063   & 0.0154$\,\pm\,$0.0010 &  0.68$\,\pm\,$0.07 & 1.46$\,\pm\,$0.133 & 0.091 & -- & -- & 0.63 & 11\,794 & 13\,343 \\
11 & NGC\,5728 & 0.0150$\,\pm\,$0.0009 &  0.50$\,\pm\,$0.18 & 2.24$\,\pm\,$0.432 & 0.216 & -- & -- & 0.41 & 26\,626 & 12\,910 \\
12 & IC\,4995  & 0.0165$\,\pm\,$0.0008 &  0.44$\,\pm\,$0.06 & 0.79$\,\pm\,$0.076 & 0.076 & -- & -- & 0.41 & 26\,966 & 13\,409 \\
13 & IC\,2560  & 0.0155$\,\pm\,$0.0011 &  0.44$\,\pm\,$0.04 & 0.62$\,\pm\,$0.059 & 0.088 & 0.22$\,\pm\,$0.05 & 0.49 & 0.28 & 53\,903 & 12\,589 \\
\cmidrule{1-11}
~~Outliers & ~~~~~~~~~~~ Average: & 0.0146$\,\pm\,$0.0020 & &  & & & & \Tosngav: & 13\,000\, & \!\!\!\!\!\!{$\pm$\,703}\,K\\ 
\cmidrule{1-11}
a & ESO\,138-G01 & 0.0366$\,\pm\,$0.0004 &  0.69$\,\pm\,$0.02 & 2.57$\,\pm\,$0.075 & 0.238 & --  & -- & 0.43 & 27\,919 & 19\,928 \\
b & Mrk\,1210    & 0.0462$\,\pm\,$0.0006 &  0.52$\,\pm\,$0.07 & 2.27$\,\pm\,$0.163 & 0.207 & -- & -- & 0.43 & 28\,404 & 23\,268 \\
c & NGC\,4507    & 0.0485$\,\pm\,$0.0004 &  0.51$\,\pm\,$0.05 & 2.75$\,\pm\,$0.127 & 0.268 & -- & -- & 0.41 & 32\,147 & 23\,918 \\
\bottomrule
    \end{tabular}
\tablefoot{
     \tablefoottext{a}{
     The list follows a descending order in the deblended {\ariv} (\arivr) doublet ratio (Col.\,9). } 
     \tablefoottext{b}{Observed blended {\arivp} (\arivtop) line ratio.}
     \tablefoottext{c}{Estimated fractional contribution \fblHe\ of the  \heiwar\ line to the observed \arivp\ \lam4711\AA\ line.}
     \tablefoottext{d}{Estimated fractional contribution \fblNe\ of the   \neiv\ \lam\lam4715\AA\ line (when detected) to the  observed \arivp\ \lam4711\AA\ line. }
     \tablefoottext{e}{Target deblended \ariv\ (\arivr) ratios after applying the blending correction(s) to the observed ratios of Col.\,4. }
      \tablefoottext{f}{Density values inferred from the deblended \ariv\ doublet ratio.}
      \tablefoottext{g}{Temperature values deduced from the \Roiii\ ratios assuming a single density \nsng. 
      }
}
\end{table*}

Visual inspection revealed significant stellar template mismatch in some objects in the wavelength range encompassing the \ariv\ doublet lines. 
We therefore ran a series of tailored fits to obtain fluxes for these weak lines.
Due to the low S/N of these lines, they were fitted with only a single Gaussian component to avoid introducing additional systematic errors. 
We first attempted to obtain \ariv\ fluxes from the 1-component fits to the B3000 spectra discussed in the paragraph above, where visual inspection was used to assess the quality of the fits. 
If they were deemed unsatisfactory, we ran a second fit on a smaller wavelength range encompassing the \ariv\ lines, including 3rd-order polynomials during the stellar continuum and emission line fits, where the order was reduced from 12 to compensate for the smaller wavelength range. 

There was a subset of nine objects for which visual inspection of the \ariv\ fits from both methods described above was not adequate. In these galaxies, the structure of the noise and/or the continuum shape underneath the \ariv\ doublet and/or the faintness of the \ariv\ lines (in particular \ariva) led to severe systematic errors in the \textsc{lzifu} fits. For these objects, we manually measured the flux of each \ariv\ line by fitting a single Gaussian to each line, and by requiring the two doublet lines to have consistent FWHM within the errors. As an additional test, we measured the fluxes simply by integrating the flux underneath each line profile after subtracting the stellar continuum fit generated by \textsc{lzifu}. The continuum level and shape were also defined separately for each individual emission line. Because in these objects the shape and level of the continuum are the main source of uncertainty, in particular due to the undulating stellar contribution, different assumptions had to be made. The reported flux errors in these affected objects account for the dispersion of the flux values from both methods.

\subsubsection{Extinction correction of the measured line fluxes}
\label{sec:extinc}

All emission line fluxes were corrected for extinction intrinsic to the AGNs using the \ha\ and \hb\ fluxes. The internal galaxy $A_V$ was computed using the \citet{Fi99} extinction curve with $R(V) = 3.1$ and assuming an intrinsic Balmer decrement $\ha/\hb$ of 3.1, as is appropriate for regions dominated by the hard ionizing spectra of AGNs~\citep{Do15,TA17}.

\begin{figure*}
\includegraphics[width=9.2cm]{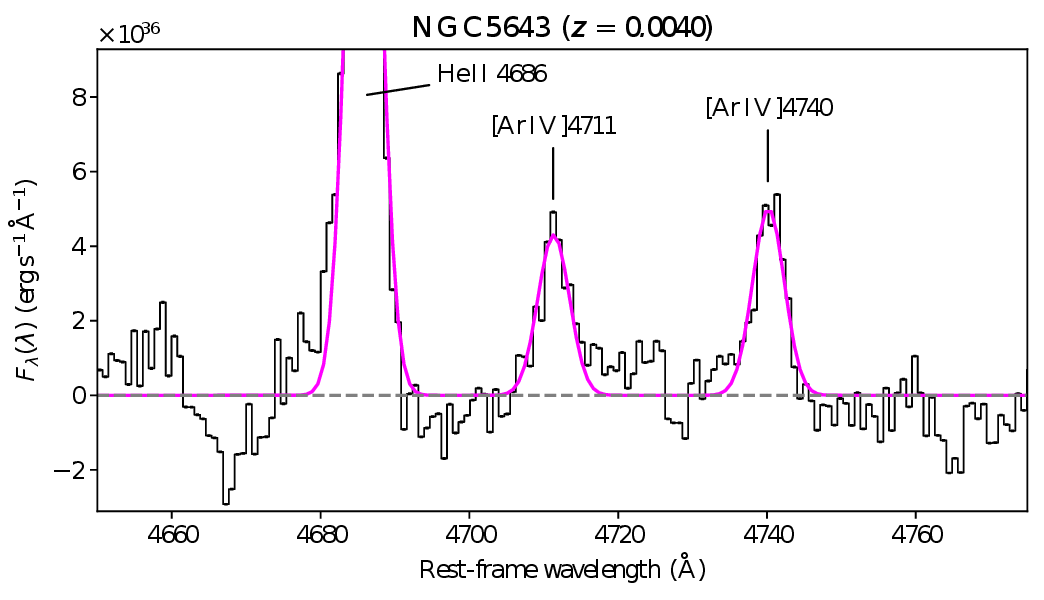} 
\includegraphics[width=9.2cm]{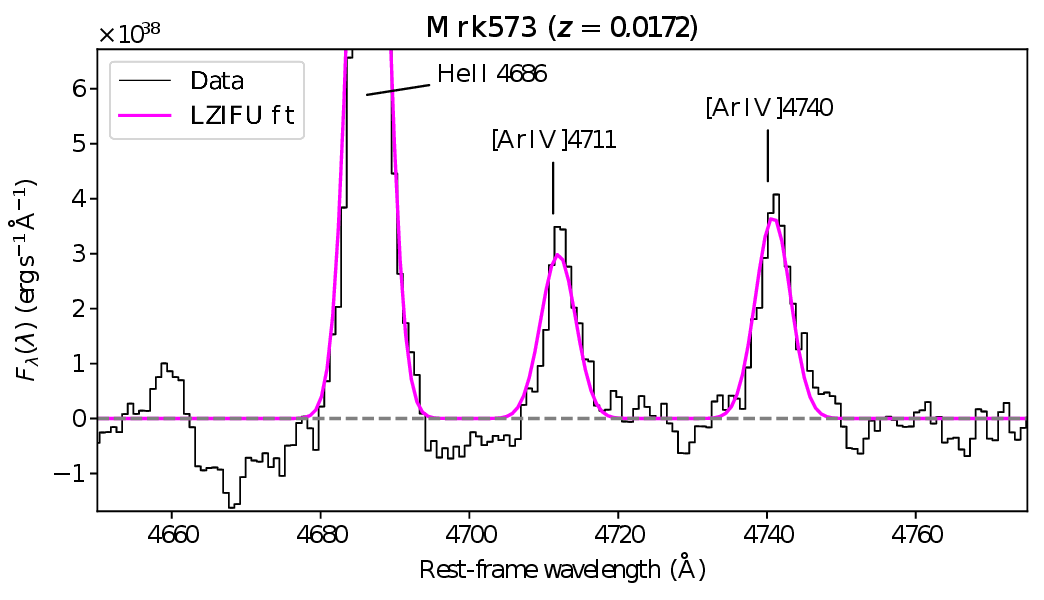}
\includegraphics[width=9.2cm]{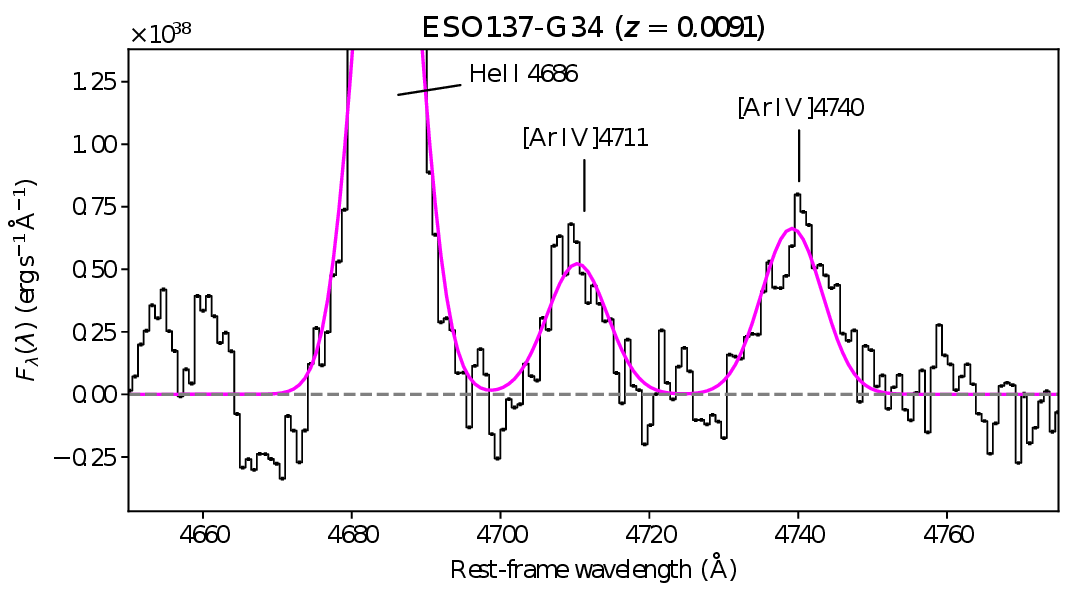}
\includegraphics[width=9.2cm]{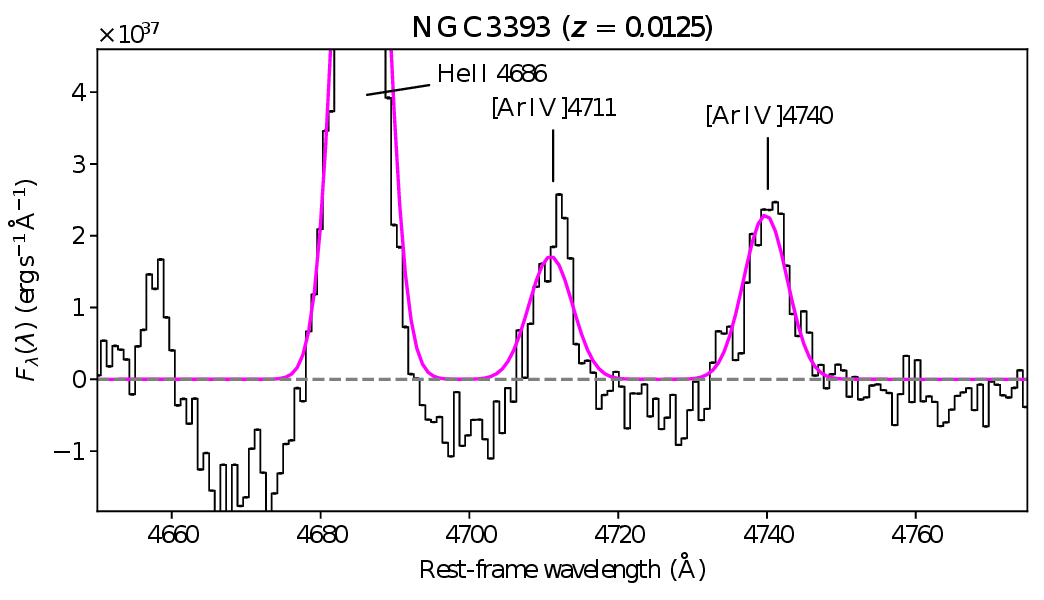}
\includegraphics[width=9.2cm]{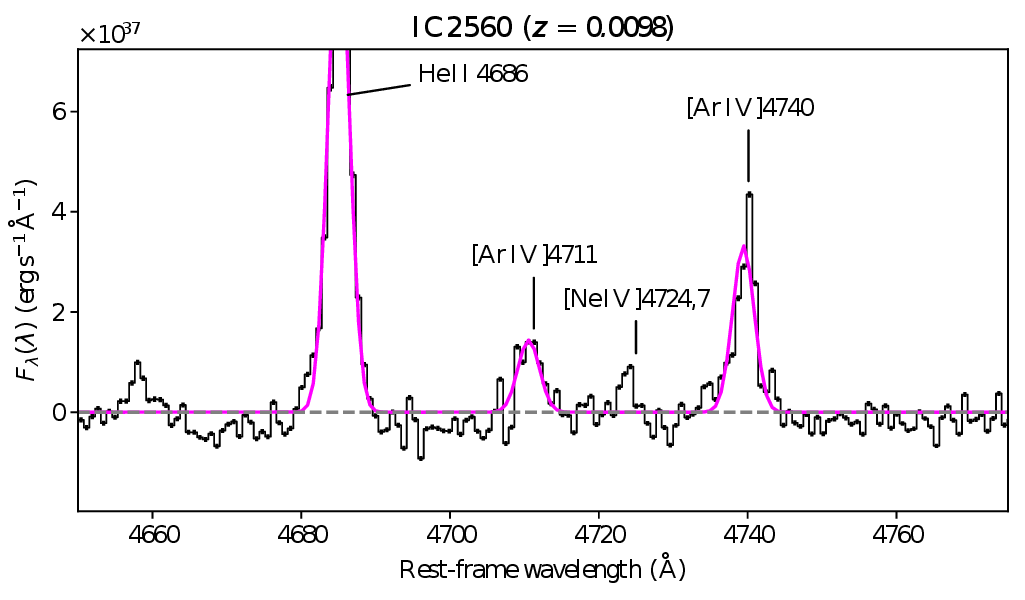}
\includegraphics[width=9.2cm]{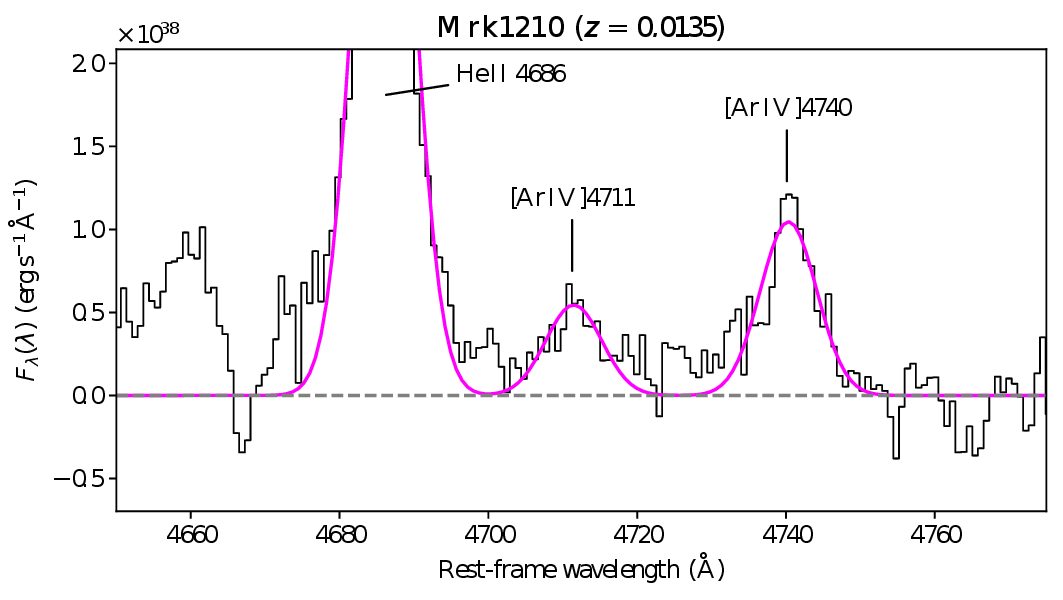}
\caption{Spectra showing the \ariv\ doublet lines of six Seyfert\,2s (black solid line) from our sample after subtraction of the stellar and AGN continuum fit. From left to right:  NGC\,5643, Mrk\,573, ESO\,137-G34, NGC\,3393, IC\,2560 and Mrk\,1210. The \textsc{lzifu} emission line fit (magenta line) is superimposed to each spectrum.
}
\label{fig:fig2}
\end{figure*}

\begin{figure}  
\centering
\includegraphics[angle=-90.0,width=\columnwidth]{figap.eps}
\caption{Pearson correlation coefficients for our sample of high excitation Seyfert\,2s from the  S7 survey. From bottom to top panel, the V band extinction in magnitudes \Av\ and the extinction corrected values of observed \arivp\ (\arivtop), \oiii\ (\oiiirb) and
\Roiii\ line ratios (Tables~\ref{tab:sam} and \ref{tab:sng})  versus the aperture $D$ (in kpc) used to extract the spectra. The  value of the Pearson correlation coefficient ($P$) is indicated in each panel. The red points identify the three outliers discussed in Sect.\,\ref{sec:outl}. } 
\label{fig:fig3}
\end{figure}

\subsubsection{Deblending  of \ariv\ \ariva\  due to weak \hei\ \lam4713\AA\ and \neiv\ \neiva\ lines }
\label{sec:deblend}

Following the dereddening of line ratios, the next step consisted of deblending the observed \arivp\ \ariva\ line$^4$, which is blended with the much weaker \heiwar\ line, as pointed out by \citet[][]{Ke19}, as well as with the \neiv\ \neiva\ line. 
We note first that, in our \textsc{lzifu} fits, we did not separately fit the weak \hei\ \lam4713\AA\ and \neiv\ \neiva\ lines, which are blended to the dominant \ariv\ \ariva\ line. The reason is that  in the majority of galaxies in our sample, the magnitude of the stellar template mismatch exceeds the amplitude of these very faint lines. Additionally, at the resolution of the B3000 spectra, given the width of the line profiles,  the \neiv\ \neiva\ lines are fully blended with the \ariv\ \ariva\ line and as a result, the fluxes of both lines would be poorly constrained were they to both be included in the fit. This is the reason for adopting the deblending procedure described in App.\,\ref{sec:ap-deblend}, which relies on measurements of the \hei/\ariv\ (\heiwarivb) ratio as well as of the  \neiv/\ariv\ (\lam\lam4725\AA/\lam4740\AA) ratio when the \neiv\ \lam\lam4725\AA\ doublet is detected. Neither deblending corrections are sensitive to the plasma temperature. 
Visual inspection of the spectra revealed that five objects (ESO\,138-G01, IC\,2560, NGC\,3281, NGC\,4939 and NGC\,5506) showed prominent \neiv\ \neivb\ emission.  In these galaxies, the \neiv\ \neivb\ fluxes (or upper limits, when the S/N was poor) were estimated by manually fitting a single Gaussian profile as done for the \ariv\ \arivb\ line in order to estimate the level of contamination of the \ariv\ \ariva\ flux by the \neiv\ \neiva\ doublet. Examples of the \textsc{lzifu} fits in the region of the \heii\ \lam4686\AA\ and \ariv\ doublet lines are shown in Fig\,\ref{fig:fig2} for six objects.

\begin{figure*}
\includegraphics[width=9.2cm]{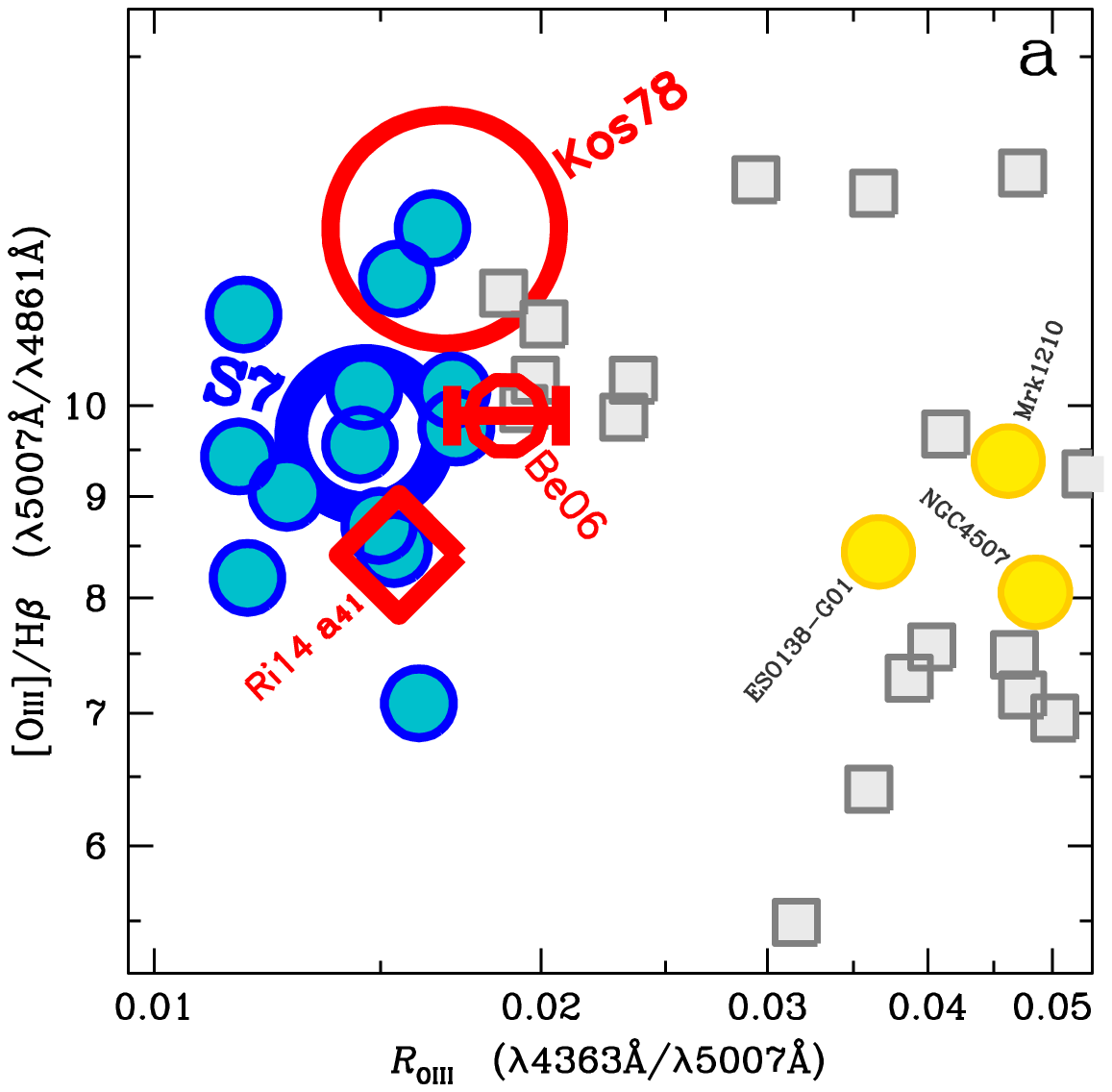}
\includegraphics[width=9.2cm]{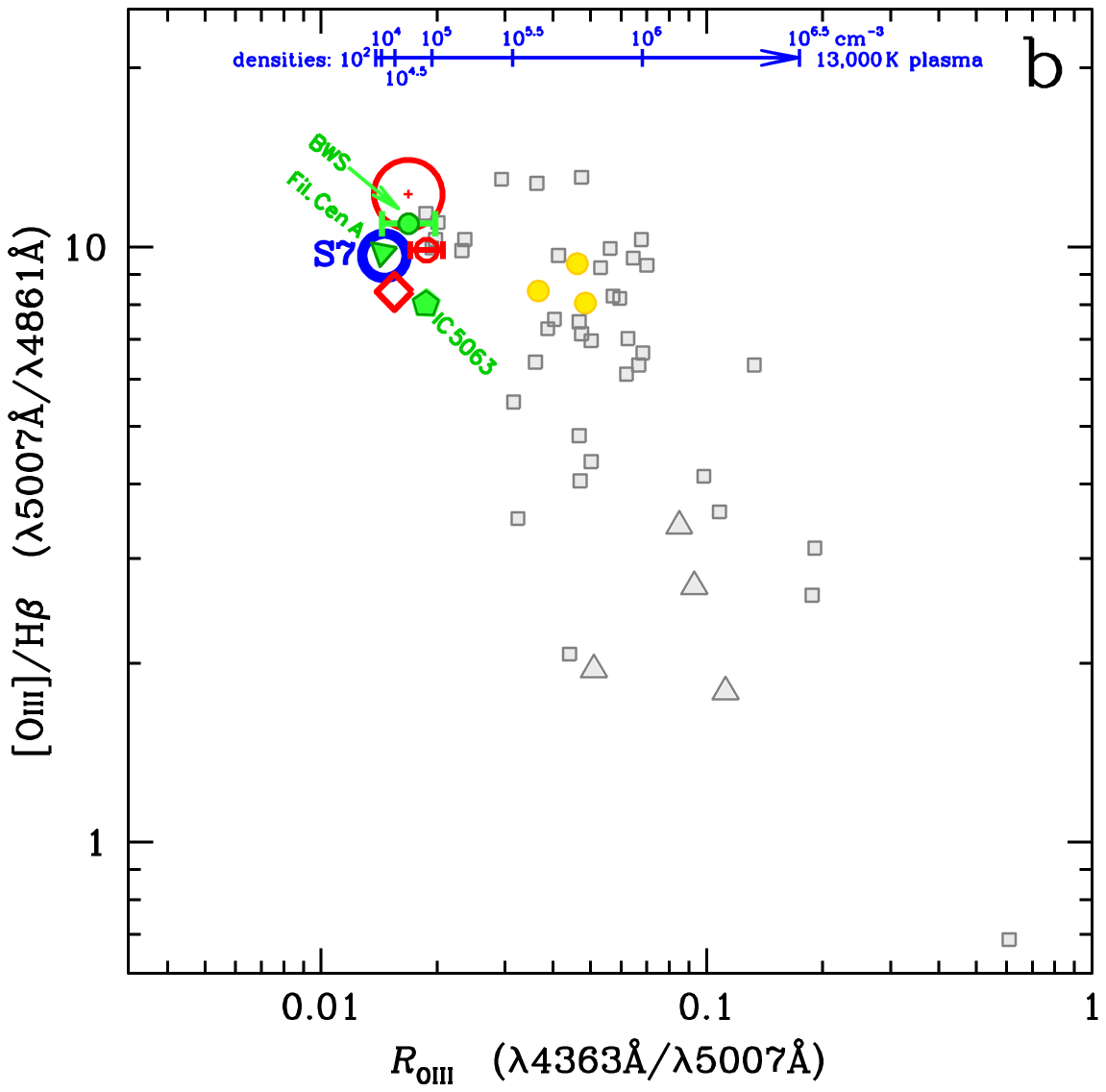}
\caption{Dereddened NLR ratios of \oiii/\hb\  vs. \Roiii. \, Left panel\textit{\,a:} blue dots represent the line ratios of 13 Seyfert\,2s from the S7 survey sample which are listed in Table\,\ref{tab:sam}. Their average \Roiii\ ratio is $0.0146\pm 0.0020$, which is represented by the navy colored circle whose radius of 0.057\,dex corresponds to the dispersion of the measurements. The yellow filled dots identifies the three outlier Seyfert\,2s: ESO\,138-G01, Mrk\,1210 and NGC\,4507 (labeled a, b, c in Table\,\ref{tab:sam}). 
Three complementary Seyfert\,2 samples  are represented by red open symbols:  1) the average of seven Seyfert\,2s from Kos78 studied by BVM (large red circle), 2) the average of four Seyfert\,2s from Be06b (small open circle with dispersion bars), 3) the high excitation Seyfert\,2 subset a41 from \citet{Ri14} (open red diamond). The light-gray open squares represent the ratios of Type\,I AGNs measured by BL05. \, \\
Right panel\textit{\,b}: a similar graph but with an expanded scale in order to cover the full range of ratios observed by BL05 among 30 quasars. The light-gray open triangles represent the four narrow-line Seyfert\,1 galaxies observed by RA00. Both BL05 and RA00 subtracted the BLR profile component from \hb\ before deriving the \oiii/\hb\ ratio. 
The light-green filled symbols represent the spatially resolved ENLR  measurements from three studies: 1) the average of two Seyfert\,2s and two NLRGs (small light-green dot) from BWS, 2) the integrated long-slit spectrum of the Seyfert\,2 IC\,5063 from Be06b (pentagon), 3) the average of seven spatially resolved optical filaments of the radio-galaxy Centaurus\,A  (light-green triangle) from Mo91. The  blue segmented arrow at the top of panel\textit{\,b:} describes the effect of collisional deexcitation on the \Roiii\ ratio for a 13\,000\,\degk\ isothermal plasma whose density is increased to successively larger values, from $10^2$ to $10^{6.5}\,$\cmc.  
 }  
\label{fig:fig4}
\end{figure*}

\subsubsection{Line ratios from the high excitation S7 Seyfert\,2 sample}\label{line_ratios}

In Table~\ref{tab:sam}, we indicate for each object the aperture that was selected (Col.\,3), the object redshift  (Col.\,4), the extinction values \Av\ (Col.\,5) inferred from the observed Balmer decrement, and finally the reddening corrected ratios of \oiii\ (\oiiirb) (Col.\,6) and \heii/\hb\ (\lam4686\AA/\lam4861\AA) (Col.\,7). 

In Table\,\ref{tab:sng}, we subsequently list the dereddened ratios of \oiii\ (\oiiir) (Col.\,3), \arivp\ (\arivtop) (Col.\,4), \RHeAr\ (\heiwarivb) (Col.\,5) and \neiv/\ariv\ (\neivb/\arivb) (Col.\,7).  The subindex $+$ refers to blending being present in the \ariv\ \lam4711\AA\ line measurement due to the unresolved \hei\ \lam4713\AA\ and \neiv\ \neiva\ lines. The adopted procedure for deblending \arivp\ (\arivtop) is described in Sect.\,\ref{sec:deblend} using the \hei\ \lam5876\AA\ and \neiv\ \neivb\ lines, respectively. This correction is relatively minor since the inferred intensity of the \hei\ (\lam4713\AA) line turns out typically to be only $\simeq 12\pm 5.3$\%  of the (deblended) \ariv\ \lam4711\AA\ line flux. Finally, the deblended \ariv\ ratios that were used to constrain the NLR densities are listed in Col.\,9  of Table\,\ref{tab:sng}. 

In Fig.\,\ref{fig:fig3} we present, as a function of the aperture $D$ (in units of kpc) projected on the sky, the distribution of the following four quantities from Tables~\ref{tab:sam} and \ref{tab:sng}:  the dust extinctions \Av\ and the three dereddened ratios of \Roiii, \oiii\ (\oiiirb) and \arivp\ (\arivtop).  The red dots correspond to the three objects identified as outliers due of their anomalous \Roiii\ ratios (see Sects.\,\ref{sec:distr} and \ref{sec:outl}). No significant correlation was found between \Av\ or the line ratios  with the aperture size, as indicated by the Pearson correlation coefficients ($P$), which are all negligible ($< 0.5$). We concluded that the use of distinct aperture sizes did not have any significant impact on the results presented in this work.

\subsection{Distribution of \Roiii\ ratios among S7 high excitation Seyfert\,2s }
\label{sec:distr}

The dereddened ratios of \oiii\ (\oiiirb) versus  \Roiii\   of our S7 survey sample are illustrated in Fig.~\ref{fig:fig4}a. Unlike quasars which extends over a wide range in \Roiii\ (light-gray open squares), our S7 sample of 13  Seyfert\,2s (blue dots) cluster around a similar \Roiii\ value. A navy colored circle (labeled S7) represents their average \Roiii\ of $0.0146\pm 0.0020$. The circle's radius  $\sigma_s = 0.057$\,dex represents the sample dispersion of the \Roiii\ ratios. The three objects labeled "outliers" in Tables\,\ref{tab:sam} and \ref{tab:sng} stand out at much higher \Roiii\ values, beyond seven times the sample dispersion $\sigma_s$. Some of their characteristics are discussed in Sect.\,\ref{sec:outl}. 

It is noteworthy that a similar clustering takes place among the seven high excitation Seyfert\,2s studied by BVM\footnote{The S7 and BVM samples have one object in common: Mrk\,573.} using  measurements from Kos78. Their average \Roiii\ is represented by the red circle centered at $\Roiii=0.0168$ and whose radius of 0.088\,dex represents the sample's RMS dispersion. 
  
Interestingly,  other samples show similar trends. For instance the average of the four Seyfert\,2s studied by \citet[][hereafter Be06b]{Be06b}, which is represented by the small red open circle with dispersion bars representing the RMS dispersion. A much larger sample of Seyfert\,2s is represented by the red open diamond which corresponds to the high excitation subset a41 extracted from the SDSS survey by {Ri14}. Unfortunately, only the \ariv\ \lam4711\AA\ line was extracted by Ri14. 

To illustrate the effect of collisional deexcitation on the  \Roiii\ ratio, the blue horizontal arrow in Fig.~\ref{fig:fig4}b represents the  \Roiii\ ratios of a 13\,000\,\degk\ isothermal plasma in which the density successively takes on values that increase from $10^2\,$\cmc\ up to $10^{6.5}\,$\cmc. For NLR densities above $10^{5}\,$\cmc, we can expect the \Roiii\ ratios to take on much higher values as is indeed observed among Type\,I AGNs from the BL05 quasar dataset, which are represented in Fig.\,\ref{fig:fig4}b by light-gray open squares. On the other hand, for the S7 Seyfert\,2s, the measured \ariv\ doublet ratios from Table\,\ref{tab:sng} show evidence that the observed NLR have densities $\ll10^{5}\,$\cmc, in which case we can trust the \Roiii\ ratio as a direct temperature estimator.

\subsection{Comparison with ENLR observations}
\label{sec:enlr}

Superposed in Fig.~\ref{fig:fig4}b are the ratios observed from the spatially resolved emission component of AGNs, the ENLR. They are represented by three light-green filled symbols which were extracted from the following samples: ($i$) the average of two Seyfert\,2s and two NLRGs (filled dot) from \citet[hereafter BWS]{BWS}, ($ii$) the long-slit observations of the Seyfert\,2 IC\,5063 by Be06b (pentagon), and ($iii$) the average of seven spatially resolved optical filaments from the radio-galaxy Centaurus\,A  (filled square) from \citet[][hereafter Mo91]{Mo91}.

A remarkable feature is the near superposition of the \Roiii\ ratios from ENLR measurements with those of the NLR of Seyfert\,2s.  Detailed modeling with photoionization calculations \citep[e.g.,][]{Ta94,Be06a, Be06b} indicated that the ENLR emission occurs at low plasma densities ($\ned < 1500\,$\cmc). The simplest interpretation for the coincidence in position in Fig.\,\ref{fig:fig4}b of the  \Roiii\ ratios from the ENLR and the NLR data sets is that in both cases the  emission  corresponds to relatively low densities in which collisional deexcitation is not significant.  Interestingly, the six quasars from the BL05 sample with the lowest values in \Roiii\ share positions relatively close to that of the NLR and ENLR measurements. We would propose that the accumulation of AGNs at a similar position on the left is most likely representing a ﬂoor AGN temperature where collisional deexcitation of \Roiii\ is not signiﬁcant.

\subsection{The three outliers}
\label{sec:outl}

The three Seyfert\,2s labeled outliers (yellow filled dots in Fig.~\ref{fig:fig4}) correspond to the galaxies ESO\,138-G01, Mrk\,1210 and NGC\,4507. Although the densities inferred from the  \ariv\ doublet are comparable to the other sample objects, their \Roiii\ values are significantly higher than the average of the other nuclei. They depart from the averaged blue dots position in Fig.~\ref{fig:fig4}a by more than seven times the cluster  dispersion in \Roiii. 
This could suggest that their plasma is much hotter ($\sim 20\,000\,$\degk), possibly as a result of fast shocks \citep{Bi85,Su03}, or alternatively could be caused by a dual-density distribution where some plasma components might possess densities above $\ga 10^6\,$\cmc\ as was considered by BVM to account for the high \Roiii\ values encountered in 3 luminous QSO\,2s where they had measurements of the \ariv\ doublet. Among the optical \oiii\ and \ariv\ lines, the only one not affected by collisional deexcitation in this case is the \oiiitw\ line since its critical density is much higher, at $4.5\times 10^7\,$\cmc. With a dual-density distribution where a very dense component is present, the integrated \ariv\ doublet ratio would be relatively unaffected by such high density component.  

Interestingly, a polarized BLR component has been detected in Mrk\,1210 \citep{Tr92,Tr95,St98} and NGC\,4507 \citep{Mo00}. They were classified as S1h by \citet{VC10,VC06}. They possibly represent borderline cases between Type\,I and II AGNs. With respect to ESO\,138-G01, the detailed analysis of \citet{ABBP} reveals a strong stratification of the emitting gas clouds that correlates with the density as well as with the presence of high ionization Fe species.  

Mrk\,1210 presents a very rich coronal line spectrum that includes [Si\,X], which is a very high excitation line not frequently detected \citep{Ma07}. ESO\,138-G01 is also a strong coronal line emitter. It was classified as a "coronal-line forest active galactic nuclei" (CLIFF AGNs), which  are characterized by strong very high ionization lines, in contrast to what is found in most AGNs \citep{CC21}. 

\citet{Ma10} studied NGC\,4507 as part of a sample of Seyferts selected on the basis of previous detection of coronal lines. It was originally classified as a Seyfert\,2  \citep{DB86}, but later reclassified as S1h (i.e., Seyfert\,1.9) by \citep{VC06} because of the presence of a weak broad \ha\ in its spectrum.

Using observed IRAS mid-infrared and far-infrared continuum fluxes at 25 and 60\mic,  \citet{MKS08} derived the spectral indices \afir\ of 98 Seyfert nuclei. The average index values reported were  $\afir=-1.5\pm 0.1$ for Seyfert\,2s and $\afir=-0.8\pm 0.1$ for Seyfert\,1s. Interestingly,  for the two outliers\footnote{To determine the  \afir\ index, the fluxes were taken from \citet{MKS08} and from NED \citep{NED}.} Mrk\,1210 and NGC\,4507, the average index is $-0.46$, indicating that the continuum emission originates from dust that is hotter than in typical Seyfert\,2s.

\section{A combined temperature-density diagnostics }
\label{sec:diag}

The utility of "diagnostic" line ratios such as \Roiii, \ariv\ (\arivr) or \sii\ (\siirr) is that they are independent of metallicity since they involve ions of the same species. 
The \ariv\ density diagnostic for instance depends on density and negligibly on temperature while the \Roiii\ diagnostic ratio depends on both except in the low density regime ($n_e \la 10^3\,$ \cmc) \citep{OST89} where it varies according to temperature. Therefore, in order to reliably measure the NLR temperature, a handle on the plasma density is required, which is our motivation for focusing on data sets where the \ariv\ doublet is observed. A concern is that while the \Roiii\ ratio remains a valid temperature indicator up to densities of $\sim\!10^8$\,\cmc, the \ariv\ doublet ratio saturates and becomes progressively insensitive to the density beyond $10^5$\,\cmc. Hence, in situations where the observed \Roiii\ ratio represents the emission from multiple density components that extend beyond $\ga 10^6\,$\cmc, the \ariv\ doublet would not detect the higher densities while the \Roiii\ ratio would increase as a result of collisional deexcitation, whether or not the temperature remains the same across the different components. 

Our goal is to determine the temperature of the observed NLR  
by combining the \Roiii\ and \ariv\  diagnostics and  requiring that the inferred temperature and density simultaneously reproduce  the dereddened \ariv\ (\arivr) and 
\Roiii\ ratios of the S7 survey sample. First, we consider in Sect.\,\ref{sec:sng} the simplest case of a single density isothermal plasma. The algorithm \OSALD\ as described in App.\,C of BVM proceeds iteratively using a nonlinear least squares fit method to determine which temperature and density are required to simultaneously reproduce the \Roiii\  and \ariv\  ratios.
Subsequently in Sect.\,\ref{sec:powl} we assume a power-law distribution of the density that extends up to the density \ncut\ which is determined iteratively using \OSALD.
  
\begin{table}
\caption{Changes in NLR temperatures assuming a power-law density distribution\tablefootmark{a}}  
 \setlength{\tabcolsep}{0.4\tabcolsep}
\label{tab:pld}
\small
\begin{tabular}{clccc}
\toprule
 (1)  & \hspace{0.225cm} (2) & (3)\tablefootmark{a}  & (4)\tablefootmark{b}  & (5)\tablefootmark{c} \\
Index & Object & \ncut  & \navr &  \delt \\ 
\#  & Name  &  \cmc &  \cmc & \degk  \\ 
  \cmidrule{1-5}
1 & NGC\,5506 & 4075 & 2428 & $-$11 \\
2 & NGC\,1320 & 10\,534 & 6228 & +3 \\
3 & NGC\,5643 & 10\,821 & 6373 & $-$11 \\
4 & Mrk\,573 & 12\,715 & 7487 & +2 \\
5 & NGC\,3281 & 13\,548 & 8022 & $-$2 \\
6 & ESO\,137-G34 & 15\,086 & 8795 & $-$6 \\
7 & NGC\,4939 & 15\,476 & 9000 & $-$19 \\
8 & NGC\,3393 & 16\,789 & 9868 & $-$19 \\
9 & NGC\,7682 & 19\,767 & 11\,590 & $-$23 \\
10 & IC\,5063 & 21\,240 & 12\,420 & $-$22 \\
11 & NGC\,5728 & 51\,920 & 30\,490 & $-$88 \\
12 & IC\,4995 & 52\,592 & 31\,200 & $-$90 \\
13 & IC\,2560 & 114\,000 & 66\,700 & $-$221 \\ 
\cmidrule{1-5}
& & & \!\!\!\!\Toplav: &\!\!\!\! 12\,961$\,\pm\,$711\,K \\ 
      \bottomrule
    \end{tabular}
\tablefoot{
\tablefoottext{a}{OSALD fits of the \Roiii\ and \ariv\ doublet ratios assuming a density distribution which extends from $10^2$\,\cmc\ up to the cut-off density \ncut. The weight applied to the emissivities follows a density power-law of index $\epsi\ = +0.6$.}
\tablefoottext{b}{Weighted average plasma density taking into account the increase of the covering factor with density.}
\tablefoottext{c}{Temperature difference between \Topla\ of the power-law density case and \Tosng\ of the single density case of Table\,\ref{tab:sng}.  }
}
\end{table}

\subsection{Temperatures inferred from the single density case}
\label{sec:sng}

In Tables\,\ref{tab:sam} and \ref{tab:sng}, the Seyfert\,2 sample follows an ascending order in the densities that were inferred from the \ariv\ doublet. These extend from 6000 up to 54\,000\,\cmc.  The average temperature for the sample of 13 Seyfert\,2s  is $\Tosngav = 13\,000\pm 703$\,\degk\ (that is, $\pm$0.023\,dex). We find remarkable that the S7 Seyfert\,2s  clustered  over such a narrow range in temperature values.  
The average temperature inferred is comparable to the value of $13\,480\pm1180$\,\degk\ derived by BVM for the seven Seyfert\,2s observed by Kos78. 

\subsection{OSALD algorithm: a power-law density distribution }
\label{sec:powl}

The algorithm\footnote{The algorithm \OSALD\ is a routine of the code \map. Both software share the same atomic database.} \OSALD\ was developed to explore the effect of collisional deexcitation on the \oiii\ and \ariv\ lines when the densities extend over a wide range of values from $10^2$\,\cmc\ up to a cut-off density, \ncut.  The intention is to implement the case of NLR densities that increase radially toward the ionizing source as was observed in ENLR studies \citep[e.g.,][]{Be06b}.

\subsubsection{Transposition to a simplified spherical geometry  }
\label{sec:trans}

The algorithm  consists in integrating the line emission measures\footnote{Defined as the line emission coefficient times the electronic density.} of an isothermal multi-density plasma (hereafter MDP) of uniform temperature \temp. The calculations can be transposed to the idealized geometry of a spherical (or conical) distribution of ionization bounded clouds whose densities $n$ decrease as $r^{-2}$. The clouds can be visualized as being radially distributed along concentric shells of negligible covering factor. The weight attributed to each plasma density component is set proportional to the covering solid angle\footnote{$\Omega(n)= A(n)/4\pi r^2$ where $A$ is the area of a shell of density $n$ exposed to the ionizing source at a distance $r$. For definiteness we set the electron density equal to that of H: $n=n_e=n_H$.} $\Omega(n)$ subtended by each plasma shell. In the case of photoionization models, such a cloud distribution would result in a constant ionization parameter \uout\ and the integrated columns $N_{\rm X_k}$ of each ion $k$ of any cloud would be to a first order constant. For the sake of simplicity, to describe $\Omega(n)$ we adopt a power law $(n/\nlow)^{\epsi}$, which extends from $\nlow\ = 10^2\,$\cmc\ up to \ncut. If we transpose this to a spherical geometry where both \uout\ and $\Omega$ are constant (i.e., $\epsi=0$), the area covered by ionization-bounded emission clouds would increase as $r^2$, thereby compensating the dilution of the ionizing flux and the density fall out (both $\propto r^{-2}$). In this case, the weight attributed by \OSALD\ to each shell is the same, otherwise  when $\epsi \ne 0$ the weight is simply proportional to $\Omega(n)$. MDP calculations are not a substitute to photoionization calculations. They serve as diagnostics that could constrain some of the many free parameters that characterize multidimensional NLR models, including the option of a nonuniform dust distribution where the opacity correlates with density, that is with $r$, as was explored by BVM. The line ratios  used as target in their diagnostics were not dereddened but were part of the simultaneous modeling of the \oiii, \ariv, \hei\ and \hi\ Balmer lines. In the current work, we only consider reddening corrected line ratios and assume a distribution of densities that extends up to a sharp cut-off. The latter is a free parameter, which is constrained  through line ratio fitting using \OSALD.

To guide us in the selection of \epsi, we followed the work of {Be06b} who determined that, for a spectral slit radially positioned along the emission line cone, the surface brightness of the spatially resolved ENLR is seen decreasing radially along the slit as $r^{\delta}$ (with $\delta < 0 $), where $r$ is the projected nuclear distance on the sky. From their \oiiiw\ and \ha\ line observations of Seyfert\,2s, Be06b derived average index values of $\delta_{[OIII]}=-2.24\pm 0.2$ and $\delta_{H\alpha}=-2.16\pm 0.2$, respectively. Let us assume that such gradient extends inward, that is, crosswise the unresolved NLR.  For our assumed spherical geometry where the \ha\ luminosity across concentric circular apertures behave as $r^{-2\epsi}$, $\epsi$ is given by $-(1+\delta)/2$. Hence the value inferred for \epsi, which describes how the covering factor varies with density,  is $+0.6$ in order that a long slit projected onto our assumed spherical geometry reproduce the $\delta_{[OIII]}$ value reported by Be06b. To derive the optimal values for the selected input parameters that would reproduce as closely as possible the \oiii\ and \ariv\ target line ratios, \OSALD\ proceeds iteratively via a nonlinear least squares fit method as described in App.\,C.5 of BVM. 

\subsubsection{Temperatures inferred from a density stratified plasma }
\label{sec:comp}

The cut-off densities \ncut\ inferred from the algorithm \OSALD\ for the case of a power-law density distribution are given in Col.\,3 of Table\,\ref{tab:pld}. These extend from 4075 up to $1.14\times 10^5\,$\cmc. What governs the integrated line ratios, however, are the mean plasma densities.  The luminosity weighted densities \navr, in Table\,\ref{tab:sng} (Col.\,4), range from 2430 up to 66\,700\,\cmc.  The average temperature for the whole sample, \Toplav, is $12\,961\pm 711$\,\degk, which is essentially the same value as the single density case of Table\,\ref{tab:sng}. The main reason is that the \Roiii\ ratio varies relatively little across the density interval covered. For instance, for the densest object, IC\,2560, the shell densities of the fitted power law cover the range $10^2$ to $1.14\times 10^5\,$\cmc\ and the \Roiii\ ratios calculated by \OSALD\ across the different shells span over the range 0.01252 to 0.01547.  

As shown by the differences in temperature  \delt\  (Col.\,5) between the values derived from \OSALD\ and those of the single density case, the temperatures \Topla\ for the first 10 objects are essentially the same as those derived assuming a single density \Tosng.   For the most part \delt\ is negligible and likely represents numerical noise from the iterative procedure, except possibly for the last three objects labeled 11 to 13.

\section{Comparison with densities inferred from outflow studies}
\label{sec:outfl}

AGNs outflows are inhomogeneous and they extend over a large range of densities and geometries. Consequently the outflow rates, gas masses and kinematics that are inferred depend strongly on how the gas densities are estimated. We now review some of the techniques used.

\subsection{The use of trans-auroral and auroral lines}
\label{sec:trau}

Evidence that the densities inferred from the red \sii\ doublet are not representative of the plasma observed in the NLR, or even the ENLR plasma, has been demonstrated in studies of outflows observed in Type\,II AGNs where the emission line profile was fitted using a central component to represent the stationary plasma and one or more velocity shifted profiles that fitted the asymmetric component, the latter being interpreted as out-flowing winds.  A technique developed by \citet{Ho11} to determine NLR densities in Type\,II AGNs made use of the trans-auroral and auroral \sii\ and  \oii\ lines. The \sii\ (\siibr) and \oii\ (\oiibr) ratios are sensitive to densities up to $10^6\,$\cmc\ and when both ratios are combined, one can simultaneously determine the plasma density and the intervening dust extinction. The densities inferred by \citet{Ro18} and \citet{Sp18} for the outflowing plasma\footnote{As measured using the "broad/intermediate" line profile fits, see nomenclature of NLR profiles defined by \citet{Ho08}.} component in ten Type\,II ULIRGs ranged from $10^3$ to $10^{4.6}\,$\cmc. For the ULIRG Pks\,B1345+12, the densities inferred by \citet{Ho11} from the intermediate and very broad NLR profile components were $10^{4.2}$ and $10^{5.5}$\cmc, respectively, while for the AGN Pks\,B1934-63 the densities inferred by \citet{Sa18} for the corresponding components were $10^{4.6}$ and $10^{5.5}$\cmc, respectively.   

\citet[hereafter Da20]{Da20} similarly used the  trans-auroral and auroral \sii\ and  \oii\ lines to evaluate the densities of eight Type\,II  objects\footnote{It included one object confirmed as Type\,1.8 and another as Type\,1.9.}. They calibrated their density diagnostic diagram using photoionization calculations that took into account differences in temperature and ionization of the regions which emit the trans-auroral and auroral lines. The densities encountered ranged from $10^{3.0}$ to $10^{3.6}$\,\cmc. The median density for the sample was $1900\,$\cmc, which is more than five times higher than the median density inferred from the \siiw\ doublet measurements from their sample. After comparing  how the red \siiw\ doublet and \oiii/\hb\ ratios varied across their respective profiles, Da20 found no evidence in their sample of an emission component that could be specifically associated to the systemic velocity (as defined by the stellar absorption lines). They commented that the multiple Gaussians from their profile fits do not represent distinct outflow components  and for this reason they use the full integrated line fluxes in  their analysis of the outflow densities.

\subsection{The ionization parameter estimation method}
\label{sec:uestim}

In objects where no reliable density diagnostic is available, \citet{Ba19b} proposed a method which they labeled "ionization parameter estimation". By comparing the  \oiii/\hb\ (\oiiirb) and \nii/\ha\ (\niirb) ratios with the predictions of photoionization calculations, one can attribute a value to the ionization parameter $U$.  The plasma density \ned\ is then derived from  the expression $\ned \simeq n_H = Q_H/U c\,4 \pi r^2_{emi}$ where $Q_H$ is the photon luminosity rate from the ionizing source, which is determined by estimating the AGN bolometric luminosity assuming a standard SED. To determine \ned, this technique is only  applicable to spatially resolved measurements as it requires knowledge of the distance \remi\ separating the point source nucleus from the position of the line measurement (along the slit or at a given IFU). However,  a novel technique to estimate \remi\ within the unresolved nuclear NLR was proposed by \citep{Ba19b}. It consisted in evaluating the mean location of the dust responsible for the mid-infrared excess emission from the outflows, which were measured using the 2-Micron All-Sky Survey \citep{Sk06}. For their sample of 234 ionized outflows in Type\,II AGNs extracted from the ALPAKA catalog \citep{Mul13}, the  mass-weighted average dust location, \rdust, was $\approx 200$\,pc $\pm 0.25$\,dex \citep[Fig.\,6 of][]{Ba19b}. The mean density that characterized their whole sample was $10^{4.5}\,$\cmc, that is more than an order of magnitude above the values inferred from the \siiw\ doublet.

The density method of \citet{Ba19b} based on the estimated ionization parameter was incorporated to the Da20 study mentioned above and applied to 11 Type\,II AGNs of their sample. For their sample which integrated the emission of the nucleus with an IFU of $1.8\farcsec \times 1.8\farcsec$, they estimated the projected size of the nucleus distance as corresponding to half the IFU size, that is 0.9\farcsec. The effect of projection was subsequently taken into account by multiplying the projected radius (in pc) by a factor of 1.4, which corresponds to a disk inclination of 45\degree. The densities they encountered extended from $10^{2.9}$ to $10^{4.8}$\,\cmc.  The median density for the whole sample is 4800\,\cmc\ while the corresponding value using the trans-auroral lines method discussed above is 1900\,\cmc. As recognized by Da20, this method is most suited to spatially resolved data, where one can obtain measurements at a known projected distance \remi\ from the AGN nucleus. Since their dataset consisted of spatially integrated nucleus measurements, Da20 recommended one should make use of the derived densities in a statistical sense rather than focusing on individual values.

\subsection{Benefits of high spatial resolution HST observations of outflows}
\label{sec:stis}

Reaching much smaller projected nucleus distances \remi\ is possible by using observations from the Space Telescope Imaging Spectrograph (STIS), as shown by  \citet{Re21}. Spatially resolved observations are essential for localizing AGN feedback and determining accurately the wind  kinematics. Observations of the \ariv\ doublet was reported by \citet{Re18a} in only one object, Mrk\,573. 
In their study, the authors combined with varying weights three different photoionization models corresponding to a high, intermediate and low $U$ values at each location along the STIS slit  observations  as well as of the DIS long-slit from the Apache Point Observatory (APO).
The line ratios with respect to  \hb\  at each radial location were generally well reproduced (their Fig.\,9) across the whole ENLR except for the two \oi\ optical lines and the two \ariv/\hb\ ratios, although it had relatively little impact on their determination of the peak outflow velocity as well as other global wind kinetics.  At only two positions along the HST-STIS slit do we suggest a distinct interpretation concerning the densities inferred, that is at $+0.05$\arcsec\ and $-0.05$\arcsec\ from the nucleus, where their intermediate $U$ model favored densities of $10^{5.1}$ and $10^{5.3}$\,\cmc\ with a relative flux weight of 50\% in both cases with respect to the high and low ionization models. Because of likely optical blurring within the 0.2\farcsec\ slit width, it is not possible in our opinion to confirm the value of \remi, on which the model densities so close to the nucleus are based. Using \OSALD, the densities we infer at the corresponding positions from the  \ariv\ doublet ratios of Re18 are 22\,600 and 4220\,\cmc, which are values significantly lower. The corresponding temperatures inferred from the \Roiii\ ratios are  16\,600 and 13\,800\,\degk, respectively\footnote{To deblend the \ariv\ doublet ratio of Mrk\,573, we used the \hei\ (\lam3889\AA/\lam4740\AA) ratio since neither the \hei\ \lam5876\AA\ nor the \lam4471\AA\ line were reported.}.  

More recently, for six Seyfert\,2s, \citet{Re22} compared the densities derived from their multicomponent photoionization models to other techniques that involve more assumptions about the gas physical conditions, but require less data and modeling \citep{Re22}. For instance, the three ionization parameter values assumed ($\log U = [-1.5, -2.0, -2.5]$) were the same at  each radial location for all six objects. A good match of the observed line ratios along the slit was obtained although the assumed \remi\ values at the position of the nucleus remained unconstrained and the caveat mentioned above by Da20 concerning the densities inferred from a spatially unresolved \remi\ value would still apply to their analysis, as acknowledged by \citet{Re21}. If we exclude those calculations positioned at the closest distance from the SMBH, the densities inferred at the other radii, among their sample of six Seyfert\,2s, all fall below $10^5\,$\cmc\ except Mrk\,3.

\section{Standard photoionization calculations }
\label{sec:calc}

Having shown that collisional deexcitation of the \oiii\ optical lines is relatively insignificant among a large fraction of Seyfert\,2s from the S7 sample, we first illustrate the  difficulty in reproducing the typical  \Roiii\ ratio observed in Seyfert\,2s and subsequently explore in Sec.\,\ref{sec:sol} possible solutions. 

\subsection{Standard input parameters for {\maphead} calculations}
\label{sec:stdpar}

Most model parameters share similar values across the calculations presented below and in Sec.\ref{sec:sol}. With respect to the plasma metallicities, it is generally accepted that gas abundances of galactic nuclei are significantly above solar values. 
The values we adopted below correspond to 2.5\,\zsol\ as in BVM, a value within the range expected for galactic nuclei of spiral galaxies as suggested by the \citet{Do14} landmark study of the Seyfert\,2  NGC\,5427 using  the Wide Field Spectrograph \citep[WiFeS:][]{Do10}. High metallicity values are shared by other observational and theoretical studies that confirm the high metallicities of Seyfert nuclei \citep{SP90,Na02,Ba08}. Our selected abundance set is twice the solar reference set of \citet{AG06}, that is, with ${\rm O/H} = 9.8\times 10^{-4}$, except for C/H and N/H which reach four times the solar values owing to secondary enrichment. We can expect the enriched metallicities of galactic nuclei to be accompanied by an increase in He abundance. We followed a suggestion from David Nicholls (private communication, ANU) of extrapolating to higher abundances the metallicity scaling formulas that \citet{Ni17} derived from local\,B stars abundance determinations. At the adopted O/H ratio, the proposed scaling formula described by Eq.\,A1 in App.\,A.c of \citet[hereafter BK23]{BK23} implies a value of He/H$\,=0.12$, which is mildly higher than the ratio of 0.103 adopted by Ri14. Our calculations were dustfree and assumed a simple slab geometry in which the plasma is radiation-bounded and exposed to ionizing radiation emitted by the accretion disk  assuming an ionization parameter defined as $\uout = {\phi_0}/{c \nout}$, where $\phi_0$ is the ionizing photon flux impinging on the photoionized slab, \nout\  the hydrogen density at the {\it face} of the slab and $c$ the speed of light. 

\begin{figure}[!t]
  \includegraphics[width=\columnwidth]{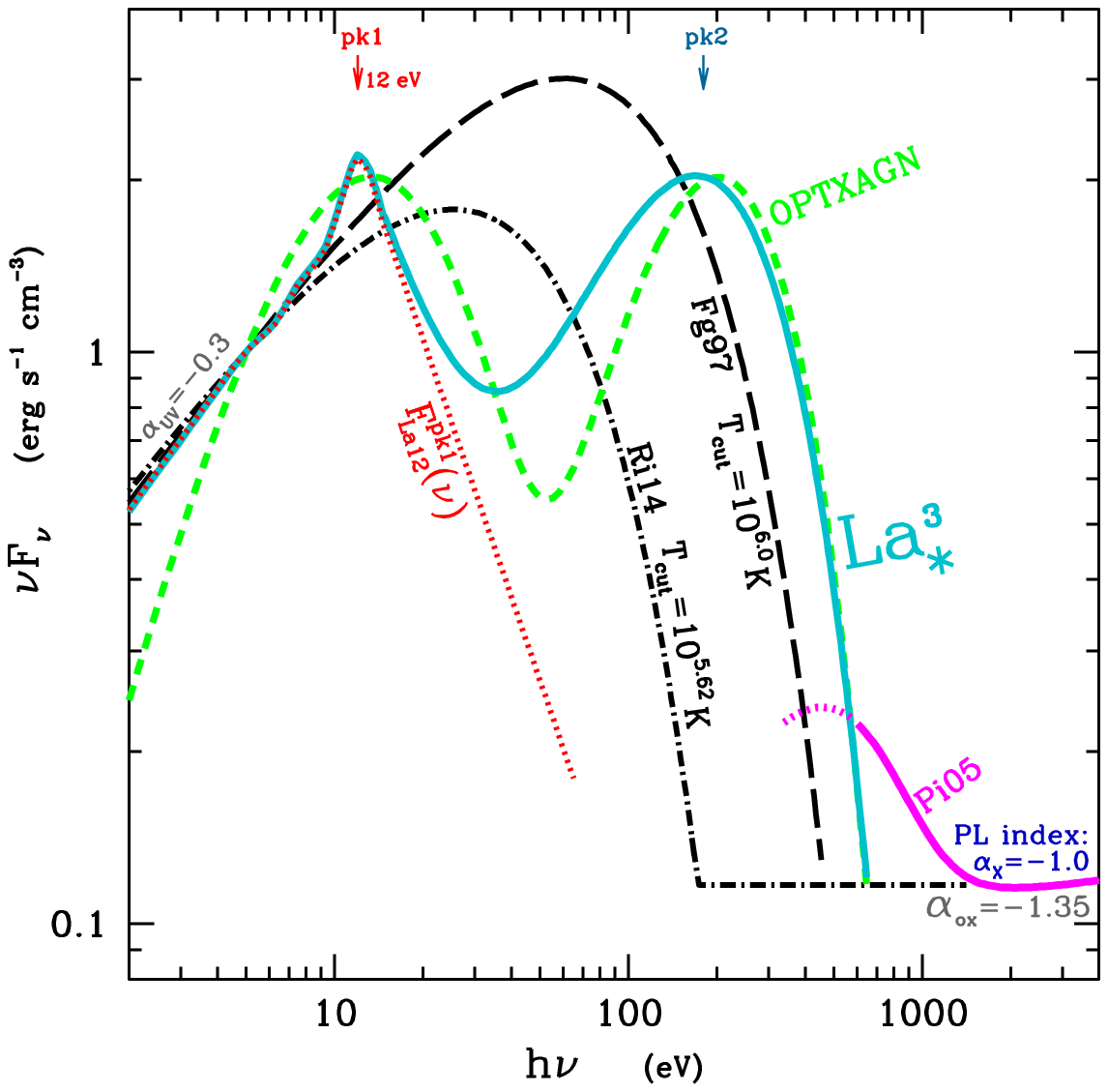}
 \caption{Ionizing SEDs in $\nu F_{\nu}$ units: 1) the SED adopted by Fg97 for their LOC calculations using $\Tcut = 10^{6.0}$\,\degk\ (black long-dash line), 2) the optimized SED of {Ri14} with $\Tcut = 10^{5.62}$\,\degk\  (black dotted short-dash line), 3) the double-bump distribution \Laasav\ with $\auv=+1.0$ (cyan continuous line) as defined in Sect.\,\ref{sec:doub}, and 4) the accretion disk model derived from  the OPTXAGN routine (light-green dashed line). All SEDs convert in the X-rays to a power law $F_{\nu} \propto \nu^{-1.0}$ with an \aox\ index of $-1.35$. The red dotted line represents a fit to the La12 UV bump at 12\,eV. The magenta line corresponds to the average soft excess component observed with XMM-Newton by Pi05.
 }  
    \label{fig:fig5}
\end{figure}

\subsection{Thermal ionizing continuum}
\label{sec:ther}

The ionizing radiation from the nucleus was assumed to originate from thermal emission by gas accreting onto a supermassive black hole. Although thermal in nature, the spectral energy distribution (SED) is broader than a blackbody since the continuum emission is considered to take place from an extended disk that covers a wide temperature range. In their photoionization models, {Fg97} and {Ri14} assumed a SED where the dominant ionizing continuum corresponds to a thermal distribution of the form  
\begin{equation}
\label{disk_eqn}
F_{\nu} \propto \nu^{\auv} \exp(-h\nu/k\Tcut) 
\end{equation} 
\noindent where  \auv\ is the low-energy slope of the "big bump", which is typically assumed to be $\auv=-0.3$, and  \Tcut\ is the temperature cut-off. The values for \Tcut\ adopted by Fg97 and Ri14 are $10^{6.0}$ and $10^{5.62}$\,\degk, respectively. Both distributions are shown in Fig.\,\ref{fig:fig5} (black dashed lines). The thermal component dominates the ionizing continuum up to the X-ray domain where a power-law of index $-1.0$ takes over. For each SED hereafter considered, we impose an \aox\ of $-1.35$.

\subsection{Standard photoionization models}
\label{sec:stdmod}

Assuming the Fg97 SED, a sequence of isochoric (i.e., constant density) photoionization models was calculated with the code \maphead\ (see updates in BK23, BVM and App.\,\ref{sec:ap-deblend} and \ref{sec:ap-C} of current paper) are shown in Fig.\,\ref{fig:fig6} (magenta line) where \uout\ increases in steps of 0.33\,dex, from 0.01 (the gray dot) up to 0.46. In each model, the same frontal\footnote{The density at the illuminated face of the photoionized gas shell. } plasma density of $\nout=10^2$\,\cmc\ is assumed. A square identifies models with $\uout=0.1$. The whole sequence predicts \Roiii\ values significantly lower than observed in the S7 survey (navy circle) or the Kos78 dataset (red circle) or the  high excitation subset a41 from Ri14 (red diamond).
The softer Ri14 SED shown in Fig\,\ref{fig:fig5} results in even lower \Roiii\ values, as shown by the yellow line sequence in Fig\,\ref{fig:fig6}. Either sequence illustrates the TE\,problem reported by various authors \citep[e.g.,][and BVM]{SB96,Be06a,VM08,Dr15}.

\section{Possible solutions to the TE\,problem}
\label{sec:sol}

Finding the cause of why model calculations predict values of \Roiii\ ratios significantly lower than observed is the next step. Except for the three outliers, where a high density NLR component is likely present, we considered that collisional deexcitation of the \oiii\ optical lines is unlikely the cause as discussed in Sect.\,\ref{sec:outfl}, at least among the current sample of 13 Seyfert\,2s where we could measure the \ariv\ doublet as well as for the seven Seyfert\,2s from the Kos78 dataset (BVM). We present below a few alternative solutions but do not rule out the existence of alternative interpretations that should eventually be explored. The calculations presented below all assume the same metallicities of 2.5\zsol\ and, except in S\,\ref{sec:doub}, the same ionizing SED of Fg97 with $\Tcut = 10^{6.0}$\,\degk.

\subsection{Double-bump continuum distributions}
\label{sec:doub}
 
As pointed out by \citet[hereafter La12]{La12}, AGNs seem to show a universal near-UV shape, reaching a maximum in $\nu F_{\nu}$  at a wavelength of around 1100\AA,  regardless of luminosity or redshift \citep[c.f.][]{ZK97,Te02,Sc04,Bi05,Sh05}.  To address the TE\,problem, {La12} explored the possibility that a population of internally cold very thick ($N_{\rm H}>10^{24}\,$\cms) dense clouds ($n \sim 10^{12}\,$\cmc) covers the accretion disk at a radius of $\sim 30\,R_s$ from the black-hole, where  $R_s$ is the Schwarzschild radius. The cloud's high velocity turbulent motions blur its line emission as well as reflect the disk emission, resulting in a double-peaked SED superposed to the reflected SED. The first peak at $\sim 1100$\AA\ represents the clouds reprocessed radiation while the second corresponds to the disk radiation reflected by the clouds, which La12 positioned at $\sim 40$\,eV. The main  advantage of this distribution is its ability to account for the "universal knee" observed at 12\,\ev\ in quasars.  As shown by BVM, however, the resulting SED did not significantly increase the temperature of the photoionized plasma. The authors argued that the second peak must be shifted to much higher energies in order to reproduce the observed \Roiii\ ratio and thereby resolve the TE\,problem.

To achieve this, BK23 proceeded as follows. First, they extracted a digitized version of the published La12 SED and to eliminate the 40\,\ev\ peak they extrapolated the declining segment of the first peak, as represented by the red dotted line in Fig.\,\ref{fig:fig5}. For the second peak, $\Fpkt(\nu)$, they adopted the formula, $\nu^{\afuv} \exp(-h\nu/k\Tcut)$ (i.e., Eq.\,\ref{disk_eqn}). All the double-bump SEDs which they explored were obtained by simply summing both distributions:
\begin{equation}
\label{bump_eqn}
   F_{\nu} = \Fpko(\nu) + R \, \; \rpk  \, \nu^{\afuv} \exp{(-h\nu/k\Tcut)}
\end{equation} 
\noindent where $R=\Fpko(\nu_{pk1})/\Fpkt(\nu_{pk1})$ is the renormalization factor which we define at $h \nu_{pk1}= 12\,$\ev, the energy where the first peak reaches its maximum in $\nu F_{\nu}$.  The position and width of the second peak depend on both parameters \afuv\ and \Tcut\ while its intensity is set by the parameter \rpk. After experimenting with different shapes and positions for the second peak, it was found that the presence of a deep valley at $\simeq 35$\,\ev\ increased the heating rate due to \hep\ photoionization, hence generating a higher plasma temperature (i.e., a higher \Roiii\ ratio). 

After comparing the plasma temperatures reached when different combinations of the parameters \Tcut, \afuv\ and \rpk\ were considered, BK23 concluded that the optimal position for the second peak is $\approx 200\,$\ev\ in order that the  \Roiii\ ratio from photoionization models reach the observed values.  Moving it to higher values was not an option as it generated an excessive flux in the soft X-rays that is not observed in Type\,II AGNs.  The height of the peak is set by the scaling parameter \rpk. For a given value of \afuv, an increase of \rpk\ or \Tcut\ favors higher values of \Roiii. However, the optimization of the SED parameters must ensure that it does not generate an excessive flux beyond 500\,eV in the soft X-rays. 


\begin{figure}[!t]
\includegraphics[width=\columnwidth]{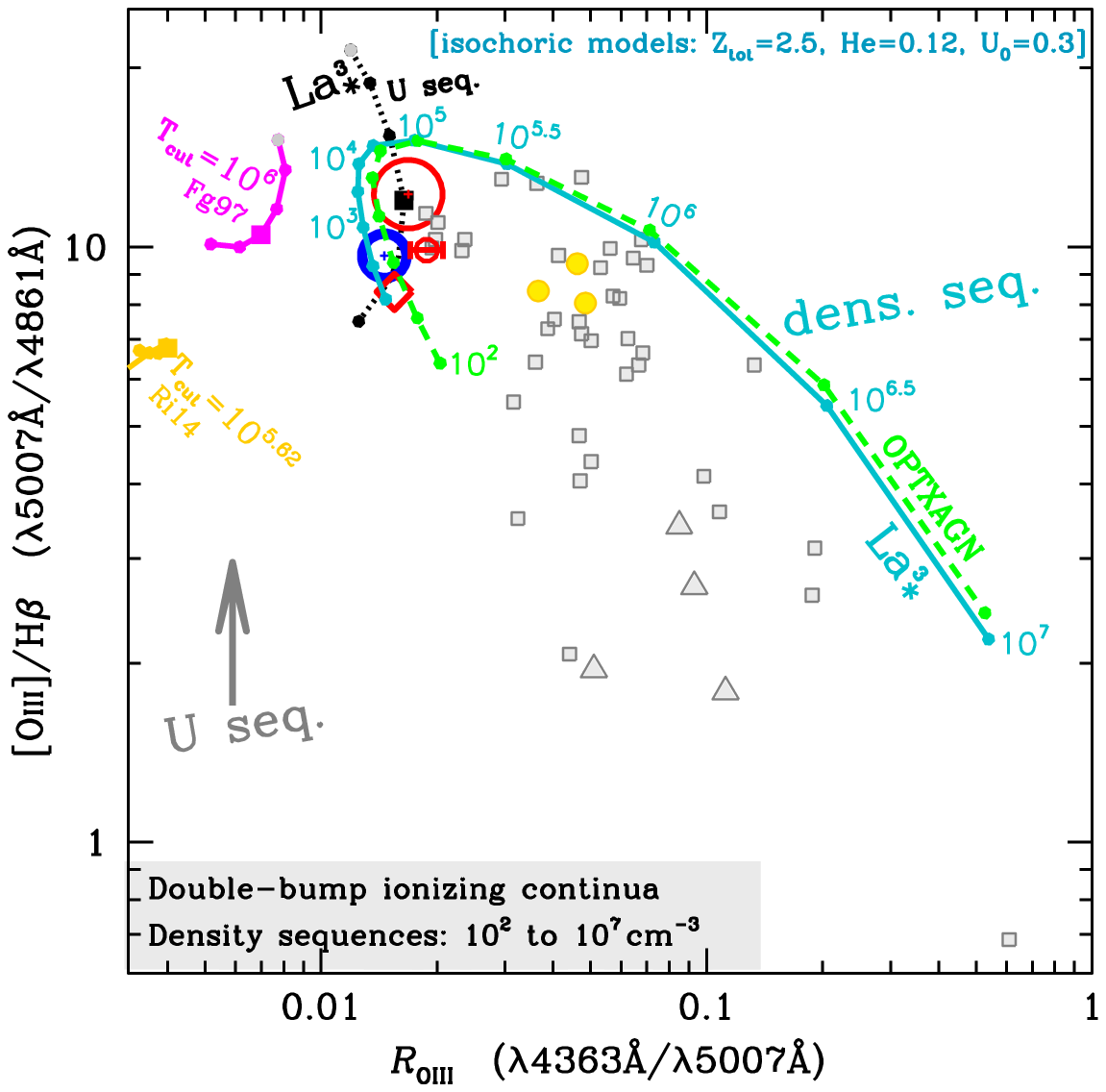}
\caption{Dereddened NLR ratios of \oiii/\hb\  vs. \Roiii\ for the same observational dataset as in Fig.\ref{fig:fig4}. 
Two ionization parameter sequences of isochoric photoionization calculations with a frontal density of $\nout=10^2$\,\cmc\ are overlaid which assume either the Fg97 (magenta) or Ri14 (yellow) SED.  A square identifies the  $\uout=0.1$ model. The third ionization parameter sequence  (black dotted line) assumed the double-peaked \Laasav\  SED  described in Sect.\,\ref{sec:doub}. Overlaid are two density sequences of isochoric photoionization models along which the density \nout\ increases in steps of 0.5\,dex, both having the same $\uout = 0.3$ value but assuming either the \Laasav\ SED (cyan solid line) or the accretion disk SED (light-green dashed line) from the OPTXAGN routine. 
} 
\label{fig:fig6}
\end{figure}

The first SED described by BK23, labeled \Laasto,  assumes the value $\afuv=+0.3$, which is similar to the standard Shakura-Sunyaev  accretion disk model\footnote{In BK23, this value was erroneously qualified as equivalent to the \auv\ assumed by Ri14 and Fg97, which is $\afuv=-0.3$.} \citep{SS73,Pr81,Ch19} with $\afuv = 1/3$. The optimized values of the other parameters were $\Tcut = 1.6 \, 10^6\,$\degk\  with a scaling factor $\rpk = 0.08$. 
Increasing \afuv\ resulted in a narrower second peak, which prevents having a conflict with the soft-X ray observational limits. The second SED proposed by BK23, \Laastd, assumed a much larger \afuv\ of $+3$. In order that the second peak occurred at essentially the same energy as in the previous SED, the parameter \Tcut\ had to be reduced  to $0.5 \, 10^{6}\,$\degk\ with a scaling parameter \rpk\ of 0.001. 

\subsubsection{Calculations using the double-bump SED \Laasav}
\label{sec:bknew}

For the current work, we opted for an intermediate case where $\afuv=+1.0$. The resulting SED is labeled \Laasav\ in Fig.\,\ref{fig:fig5} (cyan solid line). The optimized values for the other parameters are $\Tcut = 1.0 \, 10^6\,$\degk\ with a scaling factor $\rpk = 0.03$.
These parameter values ensured that the predicted flux beyond 500\,\ev\ did not exceed the soft X-rays measurements. For illustrative purposes, we show in Fig.\,\ref{fig:fig5} the average soft excess component observed with XMM-Newton (magenta line) by \citet[][hereafter Pi05]{Pi05}. It corresponds to the best-fit average of 13  quasars with $z < 0.4$ using the parameters from Table\,5 of Pi05, as described in \citet{HC07}. It has been rescaled so as to reproduce an \aox\ of $-1.35$ with respect to the first bump. We note that the dotted section below 600\,\ev\ is  speculative as it is not reliably constrained by the X-ray measurements. 

Using \map, we calculated an ionizing parameter sequence assuming a constant  density of $10^2\,$\cmc\ and the same metallicities as defined in Sect.\,\ref{sec:stdpar}. The models are represented by the black dotted line in Fig.\,\ref{fig:fig6}, which shows that the double bump SED has the potential of reproducing the observed \Roiii\ ratios. The \heii/\hb\ ratio from this sequence is 0.27, which is close to the mean value of 0.24 from the S7 sample in Table\,\ref{tab:sam}.

In order to extend our models to Type\,I AGNs (open gray symbols), we calculated a density sequence along which the density increases in steps of 0.5\,dex, from $\nout = 10^2$ up to $10^7$\,\cmc. These calculations are represented by the cyan solid line in Fig.\,\ref{fig:fig6}. We selected $\uout=0.3$ in order that the models cover the upper envelope of the quasar \oiii/\hb\ ratios. Among quasars, the vertical dispersion in the observed \oiii/\hb\ ratios is significant. It suggests a significant range in plasma excitation, which can be accounted for  using lower values of \uout\ or dual-density models as explored by BL05. 

\subsubsection{Possible origin of the double-peak SED }
\label{sec:optx}

Given that a double-peak SED can solve the \oiii\ TE\,problem, it is worth exploring how such a distribution might actually arise in the central regions of quasars. The pertinence of a second peak to describe the harder UV component is provided by studies that attempt to explain the soft X-ray excess in AGNs.  Two different processes have been proposed to explain this feature. 

The first one postulates the presence of relativistic blurred reflection by ionized plasma \citep[e.g.,][]{RF05}.  We note that this mechanism is exactly the same one as proposed by La12, producing a second peak near 40\,\ev. However, in this case, the relativistic clouds reflecting the disk emission need to be much hotter, with a high degree of ionization, in order to extend its blurred reflection all the way up to the soft X-ray band. As such, these models have the potential to explain at the same time the presence of a second peak near 200\,\ev, and the soft X-ray excess.

The second mechanism besought to explain the soft X-ray excess proposes comptonization of the accretion disk seed photons by a dual-coronal system \citep[e.g.,][]{Dn12}. In this self-consistent accretion model (referred to as OPTXAGN), the primary emission from the disk becomes partly comptonized by an optically thick warm plasma, forming the extreme EUV as well as the soft excess. An example of an OPTXAGN model is represented in Fig.\,\ref{fig:fig5} by the light-green dashed curve (see BK23 for more details).  The OPTXAGN SED can successfully reproduce the observed \Roiii\ ratio, as shown by the light-green dashed line in Fig.\,\ref{fig:fig6}, which is a density sequence of isochoric photoionization models similar to the previous \Laasav\ calculations. 
There are other sets of parameters in the OPTXAGN model that can match our double-peak SEDs, however, they require extreme accretion rates (L/L$_{Edd}\geq$1) that are not proper of Type\,II objects discussed in this work.

Despite these drawbacks, we note the striking similarity of the OPTXAGN SED with that of \Laasav, considering that they were built independently and with completely different scientific motivations. Overall, we find it remarkable that the two most popular mechanisms to explain the presence of the soft X-ray excess (blurred reflection or warm comptonization) might also produce in a natural way the double-peak SED needed to explain the observed \Roiii\ ratios.

\subsection{High ionization matter-bounded clouds}
\label{sec:mbib}

Another possibility which benefits from photoionization of \hep\ consists in high ionization matter-bounded clouds (hereafter MB) as was proposed in the modeling of the NLR and ENLR spectra by {BWS},  \citet{ST84}, \citet{VP92} and \citet{Mo91}. In their models, BWS truncated the photoionization models at a depth close to the position where most photons with energies above the ionization potential of \hep\ (54.4\,\ev) have been absorbed. They considered an ionizing continuum consisting of a power law with $F_\nu \propto \nu^{-1.3}$. In the current work, we adopt the thermal distribution of Fg97 with $\Tcut=10^6$\,\degk\ (long-dash line in Fig.\,\ref{fig:fig5}). 
Similarly to BWS, our photoionization models were truncated at a depth which is defined by the fraction \fmb\ of the ionizing photons that has been absorbed. We found that calculations with $\fmb=0.45$ can reproduce the observed \Roiii\ ratios of the S7 sample as indicated below.

\subsubsection{Radiation pressure stratified MB component}
\label{sec:strat}

Rather than assuming an isochoric or isobaric density profile for the MB component, we considered the effect of radiation pressure on the high ionization regions. As shown in \citet[][]{BWRS, Bi98, Do02}, radiation pressure exerted by the progressive absorption of the ionizing photons can induce a density gradient within the photoionized slabs. 
\citet{BWRS} has indicated that for high \uout\ values, the density averaged ionization parameter \uav\ across the photoionized slab tended asymptotically toward a constant value where the integrated line ratios to a first order became insensitive to \uout. One important parameter that must be defined is \fmb, the fraction of ionizing photons absorbed within MB clouds. Our selected value was inferred from the $\nout =10^2$\,\cmc\ MB calculation by determining at which depth  (\fmb) the integrated \Roiii\ ratio reached 0.015, the average ratio observed among the S7 Seyfert\,2 sample. We found that the truncation of the MB component should take place at $\fmb=0.45$, a similar value to that of BWS of 0.40.  

The resulting MB component is represented by the blue continuous line in Fig.\,\ref{fig:fig7} along which the frontal density \nout\ successively increases in steps of 0.5\,dex, starting at $10^2$\,\cmc.  The gray dotted line shows the calculated line ratios if the same photoionization models were ionization-bounded rather than MB. In this case, the \Roiii\ ratios are significantly lower than the S7 blue circle positioned at $10^{-1.836}$, except for the high density models owing to collisional deexcitation. 

The matter-bounded component cannot account for the lower excitation emission lines and an additional emission component is required. BWS explored the possibility of combining the emission spectra from both a matter-bounded (MB) and an ionization-bounded (IB) component. The relative proportion of both components is set by the parameter \ami. As in BWS, we  hypothesized that the radiation which photoionizes the ionization-bounded clouds corresponds to the UV radiation which escapes unabsorbed from the MB component, which lies closer to the ionizing source. 

For the IB clouds,  we assumed a frontal density larger by a factor ten (i.e., $10\times\nout$ from the MB model) and a smaller geometrical dilution factor of $5\times10^{-2}$ for the ionizing radiation that escapes the MB component. This is equivalent to having the IB clouds at a larger radial distance from the AGN by a factor of $\sqrt 20$ with respect to the MB clouds. Since the radiation pressure exerted on the IB component is negligible, we assumed the isobaric prescription.  The line ratios behavior as a function of the outer density \nout\ of the IB component is represented by the purple dashed line in Fig.\,\ref{fig:fig7}. Because the IB component generates relatively little \oiii\ emission, the integrated \Roiii\ ratio comes out primarily from the MB component. 

\subsubsection{Models that combine both the MB and IB components }
\label{sec:comb}

\begin{figure}[!t]
\includegraphics[width=\columnwidth]{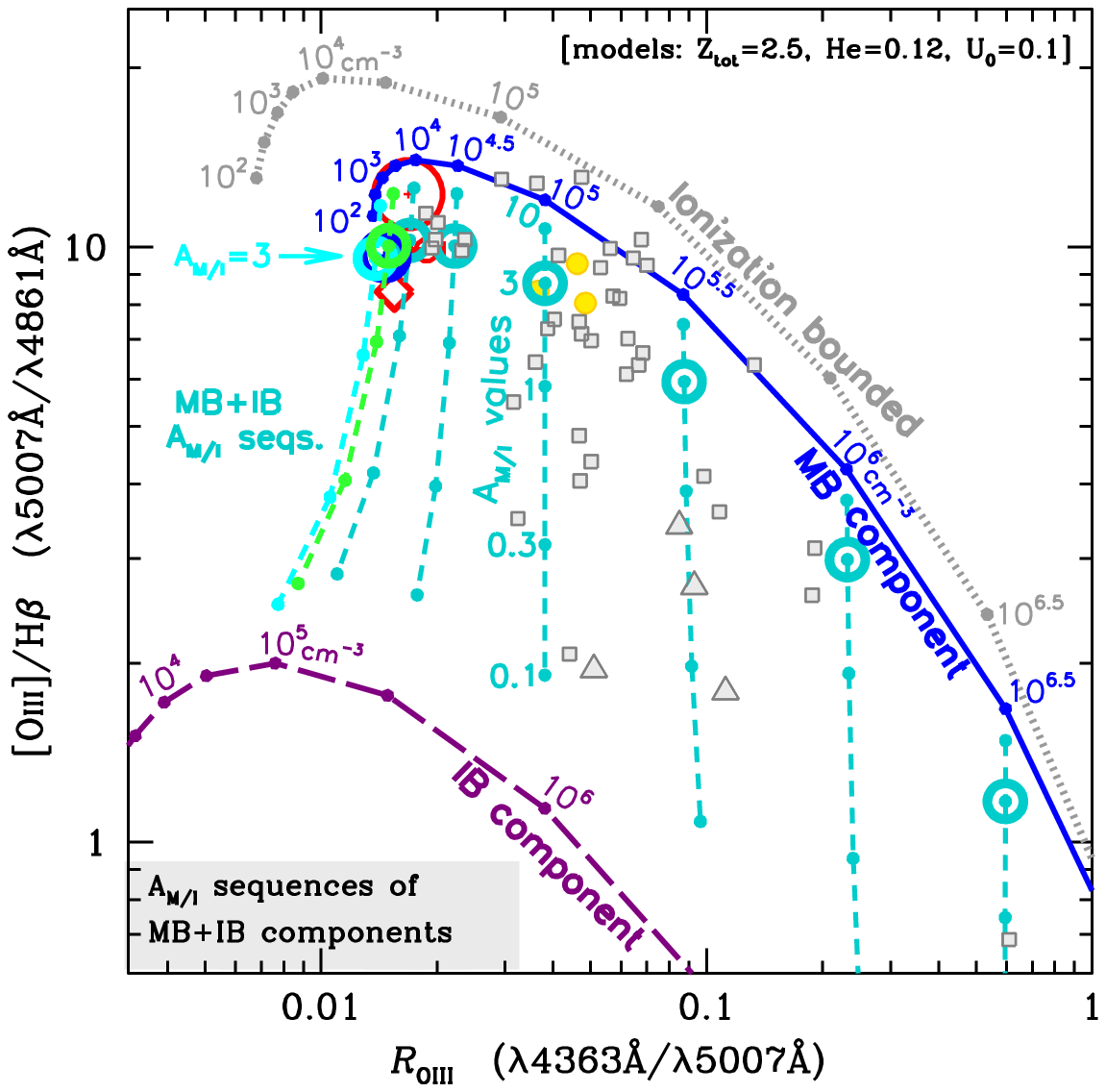}
\caption{Dereddened NLR ratios of \oiii/\hb\  vs. \Roiii\ for the same observational dataset as in Fig.\ref{fig:fig4}.
The blue solid line represent the  sequence of MB models in which the frontal densities varied from  $\nout =10^2$ up to $10^7\,$\cmc, in steps of of 0.5\,dex, assuming in all cases $\uout=0.1$. 
The gray dotted line represents photoionization models that assumed the same  parameters as the MB sequence except that the models are ionization-bounded. In both sequences the cloud internal density is stratified due to the pressure exerted by absorption of the ionizing continuum. The purple line represents the density sequence for the IB component in which the ionizing  continuum consisted of the  radiation (further diluted by a factor 20) which leaks out from the back of the MB clouds. In all calculations, the assumed ionizing continuum corresponded to  the thermal distribution of Fg97 with $\Tcut=10^6$\,\degk. The eight (near vertical) short-dash lines correspond to a linear combination of the MB model of density \nout\ with the IB model with density $10\, \nout$. The proportion between the two components was set by the parameter \ami, which successively took on the values of 10, 3, 1, 0.3 and 0.1. A circle identifies models with $\ami=3$. To avoid confusion due to the close superposition of the low density sequences, the \ami\ sequences with $\nout=10^2$ and $10^{2.5}\,$\cmc\ were left out. 
}  
\label{fig:fig7}
\end{figure}

In our approach, the relative proportion of MB and IB clouds is determined by the parameter \ami, which represents the ratio of the solid angle subtended by the MB clouds, as seen from the ionizing source, to the solid angle subtended by the IB clouds.  The line ratio \Rseq\ for any line $i$ along an \ami\ sequence is obtained using the formula:
\begin{equation}
\label{ami_eqn}
    \Rseq  =  \frac{\Rib + \Cmi \, \ami \, \Rmb}{1 + \Cmi \, \ami} 
\end{equation} 
where \Rmb\ and \Rib\ are the line ratios relative to \hb\ from the MB and IB components, respectively. By convention,  the value of the scaling parameter \Cmi\ is such that at $\ami=1$, the MB and IB components contribute in equal proportions to the \hb\ luminosity. In Fig.\,\ref{fig:fig7}, the near vertical short-dash lines illustrates how the line ratios behave as one varies the \ami\ parameter, which takes on successive values of 10, 3.0, 1.0, 0.30 and 0.10, from top to bottom. A dot identifies each \ami\ value while an open circle distinguishes the $\ami=3$ model in each sequence.  To avoid confusion, the  \ami\ sequences with $\nout=10^2$ and $10^{2.5}\,$\cmc\ were left out as they partially overlay the leftmost sequences. The two sequences with $\nout = 10^3$ (light-cyan) and $10^{3.5}\,$\cmc\ (light-green) are those that best reproduce the S7 average \Roiii\ ratio, assuming $\ami=3$ as required to fit the observed \oiii/\hb\ ratio.  The MB component in this case is reprocessing  $\simeq 75$\%  of the absorbed ionizing radiation.

The \Roiii\ ratios observed among the S7 survey Seyfert\,2s favor MB models with $\nout \la 10^{3.5}\,$\cmc\ where collisional deexcitation is not significant, which allows us to directly infer the plasma temperature. By comparison, the Type\,I AGNs measurements of BL05 and RA00 (open circles and triangles) imply much larger densities, above $10^5$\,\cmc\ in most cases. A positive aspect of the MB+IB sequences is that the integrated \ariv\ (\arivr) ratios of the $\ami=3$ models with $\nout=10^3$ and $10^{3.5}\,$\cmc\ resulted in values of 1.0 and 0.68, respectively, which is within the range of observed values among the Seyfert\,2s of Table\,\ref{tab:sng} (Col.\,9).  Furthermore, the temperature sensitive \nii\ (\niir)  ratio was calculated to be 0.013 and 0.021 for the same two models, while the mean ratio value for our Seyfert\,2 sample is $0.020\pm0.006$ according to the measurements of  \citet{TA17}. 

One objection to the proposed MB+IB calculations is the predicted  \heii/\hb\ ratio which was calculated to be $\heii/\hb\ = 0.65$ while the average observed ratio among the S7 sample of Table\,\ref{tab:sam} is $0.243\pm 0.072$. This was not an issue for the MB+IB models of BWS, essentially because the assumed ionizing SED consisted of a  power law of index ${-1.3}$ rather than the thermal SED of Fg97, which peaks at 60\ev. Another concern is the required fine tuning of some parameters such as \fmb, the density of the IB emission component or the dilution factor of the ionizing radiation reaching the IB component.


\subsection{Temperature inhomogeneities within the NLR}
\label{sec:tsqr}

In Planetary Nebulae and \hii-regions, the plasma temperatures inferred from the \oiii\ (\oiiir) ratio are significantly higher than the values inferred from recombination lines of \opp. For instance, the measurements of these lines in 20\,PNe by \citet{PP14} indicate that the temperatures inferred from the \Roiii\ ratio were  $\approx 28$\,\% higher on average than the temperatures that were inferred from  the ratio of the observed \oiiVO\ integrated multiplet\footnote{The \oiiVO\ multiplet consists of eight recombination emission lines at the wavelengths of 4651, 4673, 4638, 4462, 4696, 4642, 4676 and 4649\AA, respectively. The integrated luminosity of the whole multiplet is insensitive to the electronic density \citep{PP14,SSB14}.} 
with respect to the \oiii\ $\lambda\lambda${4959,5007\AA} lines. This phenomenon was ascribed to the existence of temperature fluctuations (that is, inhomogeneities) permeating the nebulae. Using the nomenclature of \citet{Pe67,Pe95}, the mean nebular temperature $\bar T_0$ of a nebula characterized by small temperature fluctuations \tsq\ can be defined as:
\begin{equation}
\bar T_0=\frac {\int_V n_e^2 T dV}{\int_V n_e^2 dV} \; , \label{eq:to}
\end{equation}
\noindent where $n_e$ is the electronic density, $T$ the electronic temperature and $V$ the volume over which the integration is carried out. The rms amplitude $t$ of the temperature fluctuations is given by
\begin{equation}
\tsq = \frac {\int_V n_e^2 (T - \bar T_0)^2 dV} 
{\bar T_0^2\int_V n_e^2 dV} \; . \label{eq:tsq}
\end{equation}
The values of \tsq\ derived by \citet{PP14} from their PNe sample covered the range of 0.035 to 0.128. In the case of AGN, the detection of recombination lines from \opp\ so far has not been possible owing to the weakness of the lines in relation to the bright underlying AGN continuum. For the PNe studied by \citet{PP14}, the integrated luminosity of the \oiiVO\ multiplet was rather weak, typically $< 0.2\,$\%  of the \oiii\ 5007\AA\ line. For plasma densities $<10^4$\,\cmc, the multiplet fluxes are distributed (although nonuniformly) among the eight \oiiRL\ recombination lines, which may explain why they so far have not been detected in AGNs.  As an exercise, we have explored the possibility of having temperature fluctuations within the NLR and analyzed whether they may be related to the TE\,problem described in Sect.\,\ref{sec:calc}. Assuming that the  temperature fluctuations as inferred by various authors in the case of \hii-regions \citep[e.g.,][]{PLT,ES98,RS94} were caused by an additional albeit {\it unknown} heating agent, we proceeded to quantify the energy contribution required from this heating process in order that the observed \Roiii\ values could be reproduced.  


\begin{figure}[!ht]
  \includegraphics[width=\columnwidth]{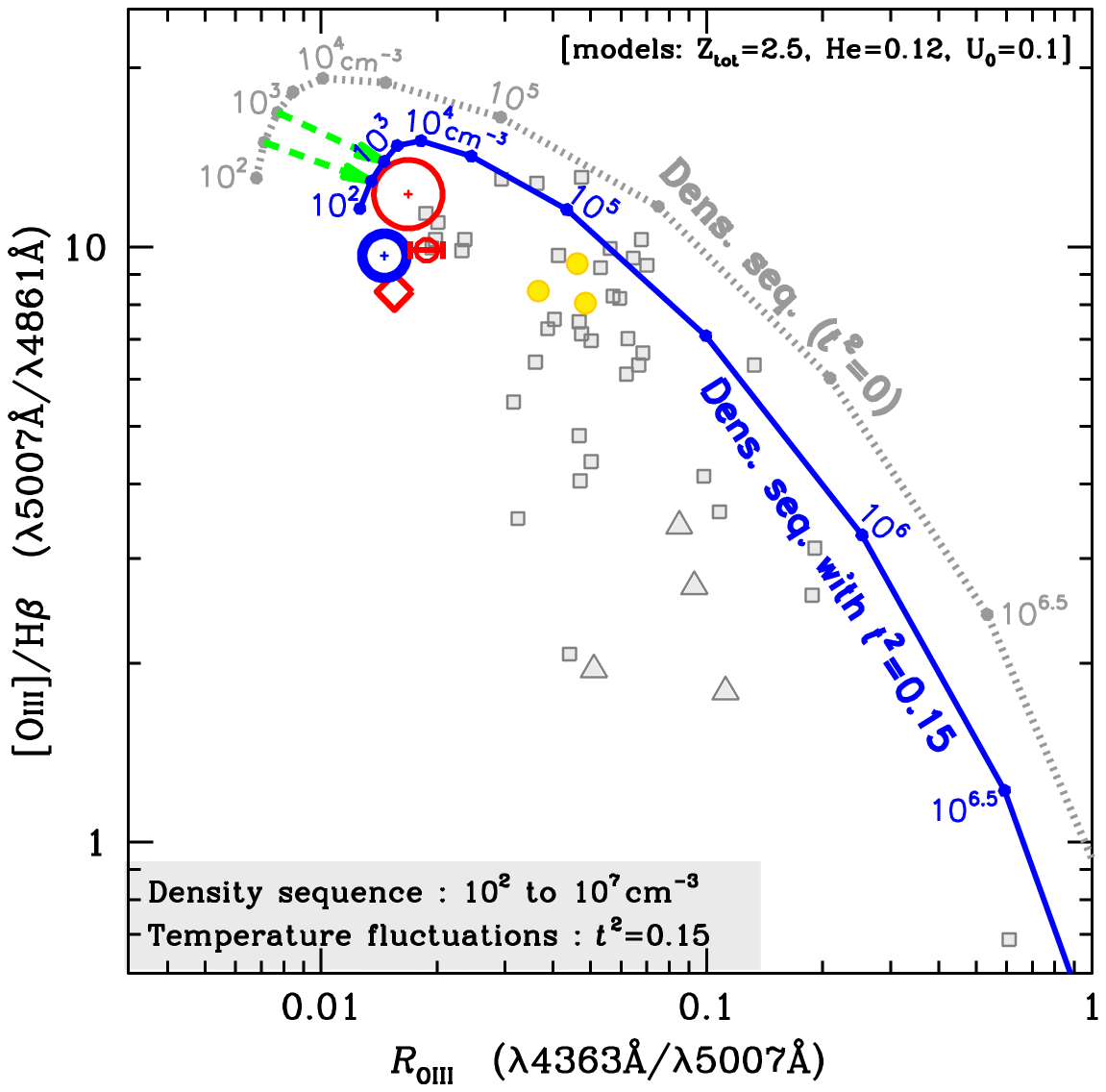}
   \caption{Dereddened NLR ratios of \oiii/\hb\  vs. \Roiii\ for the same observational dataset as in Fig.\ref{fig:fig4}.  The gray dotted line consists of radiation pressure stratified ionization-bounded models with $\uout=0.1$ assuming the Fg97 SED with $\Tcut = 10^6\,$\degk. The frontal density along the sequence varies from  $\nout =10^2$ up to $10^7\,$\cmc, in steps of of 0.5\,dex, as in Fig.\,\ref{fig:fig7}. The blue continuous line corresponds to a similar density sequence but assuming the presence of temperature fluctuations at the level of $\tsq =0.15$.  
   The light-green arrows describe the change in \Roiii\ ratios for the models with frontal densities of $\nout= 10^{2.5}$ and $10^{3.0}$\,\cmc, respectively.  }  
\label{fig:fig8}
\end{figure}

As  described in \citet[][hereafter BL00]{BL00}, the code \map\  offers the option of calculating the impact of temperature fluctuations on the emission lines from a photoionized nebula.  In photoionization calculations, it is customary to define and use at every point in the nebula a local equilibrium temperature, \teq, which satisfies the condition that the cooling by radiative processes equals the heating due to the photoelectric effect. In our hot spots scheme, by construction,  \teq\ corresponds to the temperature floor above which take place all the fluctuations.  It defines the null energy expense when calculating the extra energy emitted as a result of the small scale hot spots. The procedure consists first in defining a mean temperature \tmean\ which is derived from the computed local equilibrium temperature \teq. 
The following expression was adopted by BL00
\begin{equation}
\bar T_0 \simeq \teq [1+\gamma(\gamma-1)t^2/2]^{-1/\gamma} \; . \label{eq:tmean}
\end{equation}
\noindent The simulation of the temperature fluctuations adopted by BL00 and shown in their Fig.\,1 favored a value of $\gamma = -15$ , which is the appropriate value for any distribution of fluctuations that resembles those depicted in BL00. Subsequently, each emission line was calculated using \teq\ and  multiplied by a correction factor due to the fluctuations. In the case of recombination lines the intensity, $I_{rec}$,  of a given line is affected by a factor
\begin{equation}
I_{rec} \propto  
\bar T_0^\alpha \, [1+\alpha(\alpha-1)t^2/2] \; . \label{eq:irec}
\end{equation}
\noindent This expression, which can be used to compute individual
recombination line intensities in the presence of small fluctuations, is equivalent to calculating the intensity (which is proportional to
$T^{\alpha}$) using the effective temperature \Trec\ rather than \teq\
\begin{equation}
\Trec = \langle T^\alpha\rangle^{1/\alpha}\simeq
T_0 [1+\alpha(\alpha-1)t^2/2]^{1/\alpha} \; . \label{eq:trec}
\end{equation}

As shown by \citet[][and references therein]{Pe95}, temperature fluctuations have in general much less impact on recombination than on collisional processes which are usually governed by the exponential factor. For this reason, we adopted the simplification of considering a single value of $\alpha = -0.83$ for all recombination processes (such $\alpha$ is the appropriate value for the \hb\ line at 10000\,K). This approximation allowed us to use a single temperature \Trec\ when solving for the ionization balance of H, He and all ions of metals.

To calculate the forbidden line intensities, we solve for the population of each excited state of all ions of interest, assuming a system of five or more levels according to the ion. In the case of intercombination, fine structure and resonance lines, we treat those as simple two level systems.  More specifically, when evaluating the excitation ($\propto
T^{\beta_{ij}} \, {\rm exp}[-\Delta E_{ij}/kT] $) and deexcitation
($\propto T^{\beta_{ji}}$) rates of a given multilevel ion, each rate
$ij$ (population) or $ji$ (depopulation) is calculated using
\tmean\ (instead of \teq) and then multiplied by the appropriate 
correction factor, either
\begin{eqnarray}
{\rm cf}^{exc.}_{ij} =  
1+ \frac{t^2}{2} \Bigl[ (\beta_{ij} - 1 ) \,
\Bigl(\beta_{ij} + 2 \frac{\Delta E_{ij}}{k \tmean} \Bigr) +
\Bigl(\frac{\Delta E_{ij}}{k \tmean}\Bigr)^2
\Bigr]  \label{eq:exc}
\end{eqnarray}
\noindent in the case of excitation, or
\begin{eqnarray}
{\rm cf}^{deexc.}_{ji} = 
 1 +  \beta_{ji} \,(\beta_{ji} - 1 ) 
\frac{t^2}{2} \; , \label{eq:deexc}
\end{eqnarray}
\noindent in the case of deexcitation. These factors result in general in an enhancement of the collisional rates in the presence of temperature inhomogeneities. They are adapted from the work of \citet{Pe95} and were applied to all collisionally excited transitions. If we consider a similar distribution of  the fluctuations as assumed by BL00, the mean recombination temperature \Trec\ derived using Eq.~\ref{eq:trec} lies slightly below \tmean. Inspection of the line ratios calculated with \map\ using \tmean\ and the above correction factors confirmed that the higher is $\Delta E_{ij}$ the higher the line intensity enhancement (at constant \tsq).  It can be shown on the other hand that the far infrared lines (or any transition for which ${\rm exp}[-\Delta E_{ij}/kT] \approx 1$) are less affected by the fluctuations, similarly to the recombination lines.

Assuming a metallicity of 2.5\zsol\ and the thermal ionizing distribution of Fg97 with $\Tcut = 10^6\,$\degk\ (long-dash line in Fig.\,\ref{fig:fig5}), we explored the impact of varying the fluctuations amplitude \tsq\ until the observed \Roiii\  of 0.015 in Seyfert\,2s could be reproduced.  We found that a value of $\tsq \approx 0.15$ was required, as shown by the solid blue line in Fig.\,\ref{fig:fig8}. Interestingly, it coincides with  the upper end value inferred by \citet{MD23a} in their study of extragalactic \hii\ regions. 

We evaluated the extra energy required to generate the temperature fluctuations and found that the integrated cooling rate of the plasma with $\tsq=0.15$ was 50\% higher than when the equilibrium temperature \teq\ is assumed\footnote{This corresponds to a $\gamh$ value of 0.5 as defined by Eqn.\,12 of BL00.}, which is an uncomfortably large fraction. 
To confirm the hypothesis of small scale NLR temperature inhomogeneities would in any case require observations of the weak recombinations lines of \oiiRL\ as was achieved in PNe and \hii\ regions.  A possibility could be deep spectroscopy of the radially distant ENLR associated to radio-galaxies such as in Pks\,2152-69 \citep{Ta87} where the underlying stellar emission is weak and only a dust-reflected AGN continuum is present.

\subsection{Non-Maxwellian energy distributions}
\label{sec:kappa}

Another possibility to reach higher \Roiii\ values is to consider a $\kappa$-distribution of the electron kinetic energy rather than the standard Maxwell-Boltzmann (hereafter M-B) energy distribution. One advantage of this approach is that it does not require an additional heating mechanism as was the case with the temperature fluctuations considered in Sect.\,\ref{sec:tsqr}. What is needed, however, is a mechanism that could account for the distortion of the kinetic energies of the electrons. 

\subsubsection{Context of \textit{kappa} vs.  M-B distributions}
\label{sec:cont}
	
The assumption that the velocity distribution of the free electrons in gaseous nebulae are described by a M-B distribution is based on the electron thermalization timescale.  It is usually assumed that free electrons share their energy with the neighbouring medium so rapidly that they become thermalized before they excite any emission line. This is assumed because the energy redistribution through elastic collisions of electrons occurs faster than any other process \citep[e.g.,][]{DK18}. However, if energetic electrons are injected into the gas continually and sufficiently quickly that the particle distribution does not have time to relax to a classical equilibrium distribution, then $\kappa$-distributions may arise. 
Although $\kappa$-distributions were initially criticized for lacking a theoretical justification, \citet{Ts96} showed how these distributions can appear using entropy considerations and it has also been shown that they can arise from Tsallis’s non-extensive statistical mechanics \citep[e.g.,][]{Le02,Li09}. For a review on $\kappa$-distributions and on the mechanisms proposed to generate them in space plasmas, see \citet{Pi10}.

\citet{Ni12} proposed various mechanisms that could induce and maintain a $\kappa$-distribution within photoionized regions. These include the injection of a population of energetic electrons by acceleration mechanisms such as magnetic reconnection and development of inertial Alfvén waves, shocks, and the injection of high-energy electrons through the photoionization process. Energetic electrons can be produced by the photoionization of dust \citep{Do00}, and X-ray or EUV photoionization \citep{Sh85, Pe97}. Active galactic nuclei are regions where shocks, winds, and turbulence are known to be present \citep[e.g.,][]{Ba91, Hu08, Hu10, Za13, Si18}, and these processes may also be sufficient to accelerate particles and generate a  $\kappa$-distribution of electron energies. Several other processes, which are plausibly associated with AGN activity, may also be responsible for deviations from equilibrium, such as ionization by cosmic rays \citep{Gi05}, and dissipative turbulence internal to the gas clouds \citep{Mi19}. 

\subsubsection{Calculations assuming a $\kappa$-distribution}
\label{sec:kap}

Electron $\kappa$-distributions have been directly measured in several solar system plasmas. They are present in the solar wind \citep[e.g.,][]{Gl92, Ma97}, in the outer heliosphere and the inner heliosleath \citep[e.g.,][]{De03, He08}, in planetary magnetospheres, including magnetosheath \citep[e.g.,][]{Gl87, Kr83, Kr86, Ol68}, and in magnetospheres of planetary moons \citep[e.g.,][]{Ju02, Mo02}. 

Although it is apparent that $\kappa$-distributions are prevalent in solar system plasmas, but not until recent years has the possibility of $\kappa$-distributions in extrasolar gaseous nebulae been significantly explored. This possibility was investigated by \citet[hereafter Ni12]{Ni12}, \citet[hereafter BM12]{BM12} and \citet[]{Do13} in the context of stellar-ionized gas. 
\citet{Mo21} have recently compared photoionization models assuming a $\kappa$-distribution with the observed line ratios from a sample of 143 Type\,2 AGNs.  They found that for 98 objects, $\kappa$-distributions provide a better fit than M-B distributions.

\begin{figure}[!ht]
  \includegraphics[width=\columnwidth]{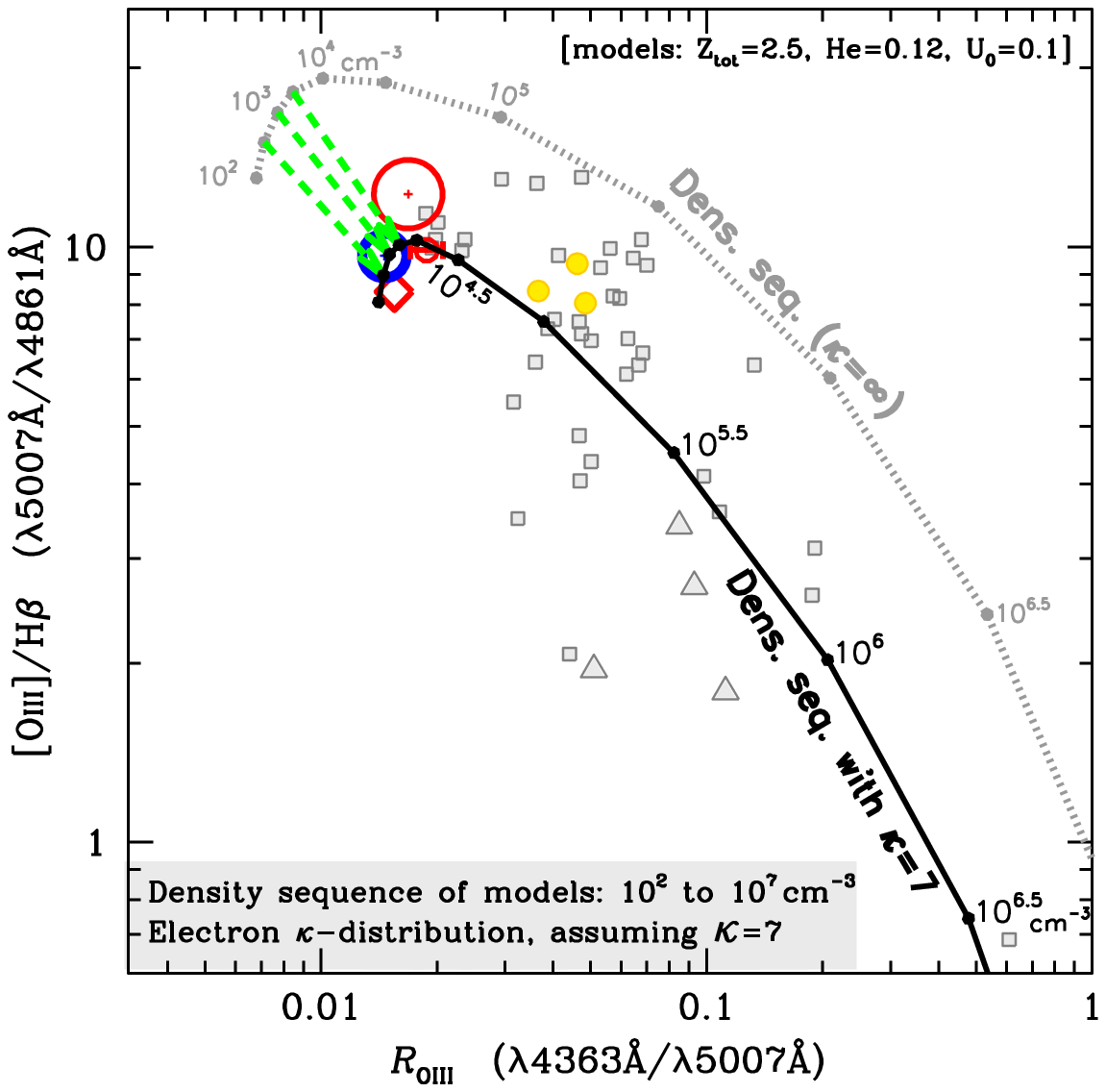}
  \caption{
Dereddened NLR ratios of \oiii/\hb\  vs. \Roiii\ for the same observational dataset as in Fig.\ref{fig:fig4}. The gray dotted line consists of radiation pressure stratified ionization-bounded models with $\uout=0.1$ assuming the Fg97 SED with $\Tcut = 10^6\,$\degk. The frontal density along the sequence varies from  $\nout =10^2$ up to $10^7\,$\cmc, in steps of of 0.5\,dex (as in Fig.\,\ref{fig:fig7}). The black line represents a similar density sequence but assuming a non-Maxwellian $\kappa$-distribution of electron kinetic energies with $\kappa = 7$. The light-green dashed arrows describe the change in \Roiii\ ratios resulting from the selected $\kappa$-distribution with $\kappa=7$.
  }  
\label{fig:fig9}
\end{figure}

In order to calculate the impact of a $\kappa$-distribution on our photoionization models, we used the analytical treatment developed by Ni12 and subsequently implemented in \MAP\ (see \citet{BM12} for details). Apart from taking into account the increased excitation rates of the atomic levels and the recombination rates of ionized species, \MAP\ also calculates the enhancement of collisional ionization that can result from $\kappa$-distributions.

Using the same metallicity as in previous models, we calculated a density sequence of ionization-bounded photoionization calculations assuming $\uout=0.1$ and the thermal distribution of Fg97 with $\Tcut = 10^6\,$\degk\  as ionizing continuum. The density sequence calculations assuming $\kappa=7$ for the electron velocity distribution (see Eq.\,1 in Ni12) is represented by the black solid line in Fig.\,\ref{fig:fig9} while the gray dotted line corresponds to the same density sequence of M-B ionization-bounded models from Fig.\,\ref{fig:fig7} (i.e., $\kappa=\infty$). In both sequences the clouds internal density is stratified due to the progressive absorption of the ionizing continuum (see Sect.\,\ref{sec:strat}).

The three light-green dashed arrows in Fig.\,\ref{fig:fig9} illustrate the change in line ratios for specific models of the M-B sequence (gray dotted line) when one assumes a $\kappa$-distribution with $\kappa=7$. The frontal densities of these models correspond to $\nout= 10^{2.5}$, $10^{3.0}$ and $10^{3.5}$\,\cmc, respectively.  They  result in an \ariv\ (\arivr) line ratio of 1.13, 0.87 and 0.54, respectively, which are representative of the observed ratios among the S7 Seyfert\,2s (Col.\,9 in Table\,\ref{tab:sng}). 

Owing to radiation pressure, the mean densities of the zones which emit the \ariv\ lines is $\approx 5 \times \nout$. Interestingly, even though the integrated H Balmer line luminosities are not affected by $\kappa$ (as they only depend on the flux of impinging ionizing photons ${\phi_0}$) the integrated columns of ionized \hyp\ in the three  models with $\kappa=7$ turned out 25\% lower because the \hyp\ recombination rates are 25\% higher than with the M-B models. 

To conclude, the models with $\kappa \simeq 7$ and densities $\nout < 10^{4}$\,\cmc\ can reproduce the observed \ariv\ and \Roiii\ ratios. Also, owing to the higher cooling efficiency  of the models that assume $\kappa$-distributions, the average plasma temperature for the  $\nout =10^3$\,\cmc\ model is actually lower, by about 1100\,\degk, than  the same model assuming a M-B distribution (i.e., $\kappa=\infty$). \citet{Mo21} has performed a more detailed comparison of the M-B distribution models with those that assume a $\kappa$-distribution. Their analysis consisted in exploring different values for the ionization parameter, the continuum spectral index, the plasma density and its metallicity. For a majority of objects from their 143 Type\,II AGNs sample, their analysis favored the $\kappa$-distribution approach.

We consider that $\kappa$-distributions have the potential of resolving the TE\,problem. We must emphasize, however, that if $\kappa$-distributions applied to the NLR, the observed \Roiii\ ratio can no more serve as temperature indicator. 
The three $\kappa=7$ models with densities similar to those  inferred from the S7 sample show an average plasma temperature of $\approx  10\,300$  rather than the previously inferred value  of $13\,000$\,\degk\  (Sect.\,\ref{sec:diag}) based on a standard M-B distribution.

\section{Summary and conclusions}
\label{sec:concl}

Following a brief review of  previous studies about the densities encountered in both the NLR and ENLR emission regions (Sect.\,\ref{sec:forbi}), we are proposing the utilization of the \ariv\ (\arivr) doublet ratio to evaluate to what extent the temperatures derived from the  \oiii\ (\oiiir) ratios of the NLR are affected by collisional deexcitation. Our Seyfert\,2 sample was originally extracted from the final data release \citep{Do15, TA17} of the Siding Spring Southern Seyfert Spectroscopic Snapshot Survey. Since our analysis centers on measurements of very weak lines, we found necessary to repeat the line extraction and line fitting procedure to ensure that the stellar spectral features surrounding the \ariv\ \arivl\ lines were properly subtracted and also to determine which aperture provided the best S/N ratio for the extracted lines (i.e., 1\farcsec or 4\farcsec). 

We extracted reliable NLR measurements of the \ariv\ doublet in 16 high excitation Seyfert\,2s of the S7 survey. The \ariv\ \lam4711\AA\ line fluxes were subsequently corrected for blending due to the weaker \hei\ \lam4713\AA\ and \neiv\ \lam\lam4715\AA\ lines. Among our sample, 13 Seyfert\,2s cluster at a similar \Roiii\ ratio of $0.0146\pm 0.0020$ while three objects labeled outliers show significantly higher values ($>0.03$). The plasma densities inferred from the \ariv\ doublet for the 13 Seyfert\,2s group span over the range $2.4 10^3$ to $5.4 10^4\,$\cmc\ in the single density case (Table\,\ref{tab:sng}). The average temperature derived from \Roiii\ for the 13 objects is $13\,000 \pm 703\,$\degk. Alternatively, when for each object a decreasing power-law density distribution is assumed that extends up to a cut-off density \ncut, the averaged values of the plasma densities that we infer for each object all turn out lower than $6.7 \times 10^4\,$\cmc\ (Table\,\ref{tab:pld}). The average temperature derived from \Roiii\ in this case is $12\,961 \pm 711\,$\degk, which is similar to the single density case. 

Assuming the thermal SED of Fg97 with  $\Tcut = 10^{6.0}$\,\degk\ (Fig.\,\ref{fig:fig5}) and a gas metallicity of 2.5\,\zsol, which is within the range expected for galactic nuclei of spiral galaxies, the photoionization models assuming low densities calculated with \map\ could not reproduce the observed \Roiii\ ratios, which we labeled the TE\,problem. We subsequently addressed this issue by exploring alternative mechanisms that could account for the observed \Roiii\ ratios while assuming plasma densities that encompassed those inferred from the \ariv\ doublet. For instance, in order to reproduce an \Roiii\ ratio close to the observed mean value of 0.015, we considered four mechanisms as follows. In Sect.\,\ref{sec:doub}, we explored the possibility of a double-bump SED where the second peak occurs near 200\,eV. One advantage of this method is that it can also account for the soft X-ray excess observed by the XMM-Newton satellite. In Sect.\,\ref{sec:mbib}, we explored the possibility of high ionization matter-bounded clouds. Although a promising solution, the calculated \heii/\hb\ ratio was much higher than observed, by a factor 2.4. In Sect.\,\ref{sec:tsqr}, following the approach of BL00, we hypothesized the existence of significant temperature inhomogeneities within the NLR clouds. To confirm this proposition would require the detection of the  \oiiVO\ multiplet. Finally, in Sect.\,\ref{sec:kappa} we considered the possibility of a $\kappa$ electron energy distribution, with $\kappa=7$. One advantage of this approach was that it did not require an additional heating mechanism although the immediate cause of the distortion of the M-B electron velocities remained indeterminate.  We note that if the $\kappa$-distribution mechanism was confirmed, the \Roiii\ ratio would no more serve as a direct temperature diagnostics. 

Independently of the origin of the TE\,problem, the fact remains that Type\,I and Type\,II AGNs reveal significant differences insofar as the optical \oiii\ lines are concerned. There is a clear  demarcation in the \oiii\ (\oiiir)  vs. \Roiii\ diagram of the Seyfert\,2’s from the S7 survey  with respect to quasars and  narrow-line Seyfert\,1 galaxies.  The most likely interpretation is that collisional deexcitation is significantly affecting the line emission of Type\,I nuclei where we have full view of the inner NLR and possibly systems at an intermediate orientation as might be the case for the few outliers of the S7 survey. The classical AGN unified model, which proposes that the NLR should be isotropic, is clearly an incomplete picture.

\begin{acknowledgements}
The S7 survey was made possible by the support of the Australian Research Council through the Discovery Project DP160103631. OLD is grateful to Fundac\~ao de Amparo \`a Pesquisa do Estado de S\~ao Paulo (FAPESP) and to Conselho Nacional de Desenvolvimento Cient\'ifico e Tecnol\'ogico (CNPq).
RAR acknowledges financial support from Conselho Nacional de Desenvolvimento Cient\'ifico e Tecnol\'ogico and Funda\c c\~ao de Amparo \`a pesquisa do Estado do Rio Grande do Sul. MVM acknowledges support from grant PID2021-124665NB-I00 by the Spanish Ministry of Science and Innovation/State Agency of Research MCIN/AEI/ 10.13039/501100011033 and by "ERDF A way of making Europe". CM acknowledges the support of UNAM/DGAPA/PAPIIT grants IN101220 and IG101223.

\end{acknowledgements}

\bibliographystyle{aa}
\bibliography{aanda.bib}


\begin{appendix}

\section{Plasma densities inferred from the  \sii\ doublet }
\label{sec:ap-sii}

Density diagnostics of the NLR typically rely on the \sii\ \siirr\ doublet ratio, which originates from the low ionization plasma. In Seyfert galaxies, the study of \citet[][]{Be06a,Be06b} for instance reveals densities of the spatially resolved ENLR that are typically $< 10^3\,$\cmc\ and in most cases both the electron density and the ionization parameter appear to be decreasing with radius. More recently, \citet{Fr18}, who used the Gemini multiObject Spectrograph Integral Field Unit (IFU) on the Gemini North Telescope, created two-dimensional electron density maps of the central part of five bright nearby Seyfert nuclei. These authors found electron density (\ned) values, inferred from the \sii\ \siirr\ line ratio, ranging from $\sim$100 up to $\sim$2500\,\cmc\ along the AGN radius. The spatial resolution quoted in their Table\,1 implies a seeing of $\sim$0.6\farcsec.  
A complementary work is that of \citet{Ka18} who generated IFU electron density maps for an AGNs sample issued from the S7 survey which consisted of seven Seyfert\,2 galaxies. The electron densities they inferred for the ENLR showed a positive gradient toward the nucleus, which was fitted with an exponential profile that initiated at a radius beyond the seeing size to avoid atmospheric blending. The density values they inferred  covered the range of 50 to 2000\,\cmc. As argued in App.\,A of BVM, for the ENLR these density values are probably not far-off from those characterizing the higher excitation lines such as \oiii. However, the situation is very different in the case of the unresolved  NLR of Type\,I AGNs as shown by BL05 (see Fig.\,\ref{fig:fig4}b) who presented  clear evidence of strong collisional deexcitation of the \Roiii\ ratio.

\section{Correcting the \ariv\ doublet from line blending}
\label{sec:ap-deblend}

The observed \arivp\ line at \ariva\ consists of a blend with the much weaker  \hei\ \lam4713\AA\ and \neiv\ \neivb\ lines. In order to properly determine the plasma densities, one must deblend the \arivp\ (\arivtoo) ratio of each object.
As in BVM, the deblended \ariv\ (\arivr) ratio is obtained by subtracting, from the measured doublet \arivp\ ratio, the fractional contribution from the \hei\ \lam4713\AA\ line (\fblHe) and the \neiv\ \lam\lam4715\AA\  line (\fblNe), that is:
$$\ariv = (1 - \fblHe - \fblNe) \times \arivp\  $$ with each blending correction  evaluated as follows.  

\subsection{Blending of the \hei\ \lam4713\AA\ line}
\label{sec:ap-debhe}

One characteristic of ratios involving recombination lines of the same ion is their limited sensitivity to either temperature or density. A valid prediction of the \hei\ \lam4713\AA\ line intensity can therefore be derived from the measurement of the \hei\ (\lam5876\AA\ line) (alternatively, one can use the \hei\ \lam4471\AA\ line). Our deblending procedure is as follows. First, we derive the Case\,B \hei\ (\lam4713\AA/\lam5876\AA) ratio, which we label \Rhei, via interpolation of the emissivities from the supplemental Table of \citet{Po13}. For a  $10^4\,$\cmc\ plasma at a temperature of 12\,000\,K, the \Rhei\ ratio turns out to be only 4.78\%  of \hei\ 5876\AA. Temperature variations of $\pm 2000$\,\degk\ would cause a change in this ratio of $^{+7.95}_{-11.5}$\%, respectively, while adopting density values in the interval 10$^2$ to 10$^5\,$\cmc\ would result in \Rhei\ ratios between   0.0429 and 0.0489, respectively.
Second, by defining \RHeAr\ as the observed \hei/\ariv\ (\heiwarivb) ratio, the product $(\Rhei \times \RHeAr)$ defines our estimate of \fblHe, which is the contribution from \hei\ \lam4713\AA\ blending to the "measured" {\arivp} doublet ratio. The \fblHe\ values for the S7 survey assuming a single density are shown in Col.\,6 of Table\,\ref{tab:sng}.

\subsection{Blending of the \neiv\ \lam\lam4715\AA\ line.}
\label{sec:ap-debne}

The \neiv\ optical lines consist of a quadruplet at \lam4714.36, \lam4715.80, \lam4724.15 and \lam4726.62\AA, respectively \citep{GR15}. To simplify the notation, we refer to the quadruplet as consisting of two doublets: the observed \neiv\  \neivb\ doublet and the potentially blended \neiv\ \neiva\ doublet. At typical NLR densities and temperatures, the observed \neiv\ \neivb\ doublet 
turns out to be $\approx 35$\% brighter than the blended \neiv\ \neiva\ doublet. The dependence of this ratio on temperature and densities is relatively minor. Within the S7 survey, we could detect the \neiva\ doublet only in a few objects. To correct the measured \arivp\ ratio from \neiv\ blending, we first calculate at the appropriate density and temperature the \neiv\  (\neiva/\neivb) ratio, which we label \Rneiv.  By defining \RNeAr\ as the observed \neiv/\ariv\ (\lam\lam4725\AA/\lam4740\AA) ratio, the product ($\Rneiv \times  \RNeAr$) defines our estimate of \fblNe, which is the contribution from \neiv\ blending to the "measured" \arivp\ doublet ratio.

\section{Recent updates to the code \maphead} 
\label{sec:ap-C} 
 
We incorporated the following recombination emission line processes to the  version \ifla\ of \MAP.
\subsection{Emission resulting from recombination of \npp, \oppp, \opp\ and \sppp\ to metastable levels}
\label{sec:ap-Ca}

Based on the work of \citet{PPB91}, the recombination rates from \npp, \oppp\ and \opp\ to the corresponding metastable levels \nso\ and \ndd\ of \nii\ and \oiii\ and levels \dpx\ and \ddx\ of \oii, respectively, have been incorporated in the calculation of the corresponding emission line intensities. In the case of the \nso\ level of \oiii, we added the missing contribution from dielectronic recombination (Christophe Morisset, private communication). As for the \spp\ and \sppp\ ions, we estimated their recombination rates to metastable levels by extrapolation from the \opp\ and \oppp\ ions as follows: we assumed that the fraction, \ram, of the total recombination rate ($\alpha_{\rm SII}^{rec}$ or $\alpha_{\rm SIII}^{rec}$), which populates metastable levels of Sulfur is the same fraction as found for Oxygen. For instance, for a 10\,000\,K plasma this fraction, \ramoiii, in the case of level \nso\ of  \oiiix\ (causing emission of the \oiiitw\ line), is $2.2$\% of $\alpha_{\rm OIV}^{rec}$.

\end{appendix}

\end{document}